\documentclass[manuscript,screen]{acmart}

\usepackage{amssymb}
\usepackage{amsmath}
\usepackage{graphicx}
\usepackage{todonotes}

\DeclareMathOperator*{\argmax}{arg\,max}
\DeclareMathOperator*{\argmin}{arg\,min}

\AtBeginDocument{%
  \providecommand\BibTeX{{%
    \normalfont B\kern-0.5em{\scshape i\kern-0.25em b}\kern-0.8em\TeX}}}

\setcopyright{acmcopyright}
\copyrightyear{2020}
\acmYear{2020}
\acmDOI{10.1145/1122445.1122456}

\begin{document}

\title{SCARL: Side-Channel Analysis with Reinforcement Learning on the Ascon Authenticated Cipher}

\author{Keyvan Ramezanpour}
\email{rkeyvan8@vt.edu}
\orcid{0003-3552-1220}
\affiliation{%
  \institution{Virginia Tech}
  \city{Blacksburg}
  \state{Virginia}
  \postcode{24061}
  \country{USA}
}
\author{Paul Ampadu}
\email{ampadu@vt.edu}
\affiliation{%
  \institution{Virginia Tech}
  \city{Blacksburg}
  \state{Virginia}
  \postcode{24061}
  \country{USA}
}
\author{William Diehl}
\email{wdiehl@vt.edu}
\affiliation{%
  \institution{Virginia Tech}
  \city{Blacksburg}
  \state{Virginia}
  \postcode{24061}
  \country{USA}
}

\renewcommand{\shortauthors}{K. Ramezanpour, P. Amapdu and W. Diehl}

\begin{abstract}
  Existing side-channel analysis techniques require a leakage model, in the form of a prior knowledge or a set of training data, to establish a relationship between the secret data and the measurements. We introduce side-channel analysis with reinforcement learning (SCARL) capable of extracting data-dependent features of the measurements in an unsupervised learning approach without requiring a prior knowledge on the leakage model. SCARL consists of an auto-encoder to encode the information of power measurements into an internal representation, and a reinforcement learning algorithm to extract information about the secret data. We employ a reinforcement learning algorithm with actor-critic networks, to identify the proper leakage model that results in maximum inter-cluster separation of the auto-encoder representation. SCARL assumes that the lower order components of a generic non-linear leakage model have larger contribution to the leakage of sensitive data. On a lightweight implementation of the Ascon authenticated cipher on the Artix-7 FPGA, SCARL is able to recover the secret key using 24K power traces during the key insertion, or Initialization Stage, of the cipher. We also demonstrate that classical techniques such as DPA and CPA fail to identify the correct key using traditional linear leakage models and more than 40K power traces.
\end{abstract}



\maketitle

\section{Introduction}
Physical exposure of Internet-of-Things (IoT) and mobile devices in unprotected environments enables the class of physical attacks, called side-channel analysis, that compromise the security of computing systems when an attacker has access to the hardware platform. While standardized cryptographic algorithms are generally secure against classical cryptanalysis, the run-time signatures of devices executing an algorithm can expose information about the processed secret data. Side-channel analysis (SCA) refers to attack techniques that attempt to recover the secret key of cryptographic algorithms by inspecting the signatures of a hardware platform. The most common signatures of hardware exploited in SCA include the power consumption \cite{mahanta2015power, luo2018power} and electromagnetic (EM) radiations of the device processing the secret data \cite{longo2015soc}.

Side-channel analysis techniques that use the power consumption of the hardware under attack to recover the secret key are often called power analysis (PA). A typical power analysis technique requires a \textit{leakage model} and proper \textit{statistics} to measure the goodness of fit into the model. The leakage model establishes a mathematical relationship between the power consumption of the device and the processed secret data. In a PA attack, the secret data is calculated from the input and/or output of the algorithm, e.g. plaintext and/or ciphertext in an encryption/decryption algorithm, and a subset of the secret key. Using the statistics, the attacker measures how well the relationship between the power measurements and the calculated secret data fits into the leakage model, for every possible value of the key subset. The correct key exhibits the highest rank according to the statistics.

Traditional PA techniques can be divided into two categories of \textit{model-based} and \textit{profiling} attacks, depending on how much prior knowledge on the leakage model is required. Model-based techniques assume a full knowledge of the attacker on the leakage model. These techniques differ in the statistics they employ for evaluating the goodness of fit into the model. Differential power analysis (DPA) \cite{kocher1999differential} uses clustering as a goodness of fit, in which the difference between the mean of power traces in two clusters is used as the statistical relation to rank the key candidates \cite{kocher1999differential}. Alternative statistics include Pearson's correlation coefficient in correlation power analysis (CPA) \cite{chakraborty2017correlation}, mutual information analysis (MIA) \cite{whitnall2011exploration} and Kolmogorov–Smirnov (KS) statistics \cite{gierlichs2008mutual}.

The most common traditional leakage model is the Hamming weight (Hw) of the binary data in which the power consumption of a logic block is correlated with the Hw of the computation result \cite{alioto2009leakage, peeters2007power}. Further, Hamming Distance (HD) between the initial and final values of memory elements is popular in modeling the power consumption corresponding to memory transitions, e.g. in registers of microprocessors \cite{heuser2012revealing}.
The toggling activity, or switching glitches, of internal nodes in the gate-level implementation of logic functions, which depends on the processed data, constitute alternative leakage models \cite{ sadhukhan2019count}.

Profiling PA attacks address the issue of model uncertainty by estimating the proper leakage model using a training, or reference, set. The reference set consists of pairs of power measurements and the corresponding known secret data collected from a hardware platform identical to the device under attack. Template attack (TA) \cite{chari2002template} is a classical example of a profiling technique in which a multivariate Gaussian distribution is fitted on the power traces in the reference set. The parameters of the distribution function, i.e. the mean and covariance matrix, depends on the secret data. Maximum likelihood (ML) or maximum a posteriori (MAP) statistics are used to evaluate the goodness of fit of the power measurements from the device under attack.

Machine learning techniques have been widely used in profiling attacks to estimate the proper leakage model and establish goodness of fit statistics. Support vector regression (SVR) \cite{jap2015support} and multi-layer perceptron (MLP) neural networks have been employed \cite{yang2011back} to develop a generic non-linear model. A partition-based approach, using k-means clustering, is also used in \cite{whitnall2015robust} to identify a set of values of the secret data that result in similar power traces. Given the extracted leakage model, classical statistics, such as correlation coefficient, mutual information or ML/MAP metrics can be used as the rank statistics. Alternatively, machine learning techniques including support vector machine (SVM) \cite{bartkewitz2012efficient}, decision tree (DT) or random forest (RF) \cite{banciu2015reliable, lerman2011side}, have been employed to evaluate the goodness of fit.

An emerging class of profiling SCA techniques is based on \textit{supervised} deep learning. In these techniques, the leakage estimation and goodness of fit evaluation are integrated into a single algorithm based on a deep neural network (DNN). During training, the neural network extracts the most relevant features from the power traces that correspond to the secret data. The feature extraction is equivalent to leakage model estimation. The loss function of the neural network, e.g. the cross-entropy loss, serves as the goodness of fit statistic. The problem of deep learning based SCA (DL-SCA) then reduces to training a DNN, with a proper architecture, that results in the minimum loss during inference, or test. Convolutional neural networks (CNN), long short-term memory (LSTM) networks, stacked autoencoders and deep MLP networks have been investigated in \cite{maghrebi2016breaking, wang2019diversity, zhang2019multi}, which show improved performance over profiling attacks using classical machine learning techniques.

Supervised learning SCA attacks are powerful if proper a training set, collected from a hardware platform identical to the device under attack, is available. This is a major limitation of supervised learning techniques. An alternative approach is using \textit{unsupervised} learning to extract the leakage model from the measurements. An SCA attack based on unsupervised learning and sensitivity analysis is introduced in \cite{ramezanpour2020scaul}. The leakage model, either in the form a prior knowledge or a training set, constitutes side information required in existing SCA techniques to retrieve information about the secret data. Unsupervised learning SCA attempts to recover the secret data based on the assumption that all information about the secret is included in the measurements, given the input and/or output of the algorithm. This assumption has significant implications for the security of hardware platforms; the access to the input or output of the algorithm with the corresponding power measurements is all that is required to recover the secret key. This assumption does not hold for the supervised learning SCA.

In this paper, we introduce \textit{side-channel analysis with reinforcement learning (SCARL)} that recovers the secret data from power measurements without requiring a leakage model or training set. 
SCARL assumes that the lower order terms in a non-linear model have the main contribution to the leakage of the secret data. We introduce a regularized maximum likelihood autoencoder, with LSTM neural networks, that extracts maximum information from the power measurements. We employ reinforcement learning, using actor-critic networks, to divide the measurements into two clusters based on pronounced features of the autoencoder, i.e. the features that exhibit the highest inter-cluster difference. Using only the lower order terms in a generic non-linear leakage model, estimated from the clustered measurements, the correct key shows the highest inter-cluster difference of at least one of the pronounced features.

Recent standardization processes, such as the U.S. National Institute of Standards and Technology (NIST) Lightweight Cryptography (LWC) Standardization Process, require assessment of the security of hardware implementations of cryptographic algorithms against side-channel analysis. The NIST competition is currently evaluating the security of authenticated ciphers and hash functions for future lightweight standards. The Ascon authenticated cipher is selected for the second round of the competition and was introduced during the Competition for Authenticated Encryption: Security, Applicability and Robustness (CAESAR) as the first choice for the lightweight use case. We demonstrate the success of SCARL in recovering the 128-bit secret key of Ascon using power measurements during S-box computations in the first round of the key insertion, or Initialization Stage, from a register transfer level (RTL) hardware implementation on an Artix-7 FPGA.

Our contributions in this work include: 1) Demonstration of the ability of reinforcement learning, in the context of SCA, to extract secret data in a self-supervised approach; 2) Introduction of a regularized maximum-likelihood LSTM autoencoder, and demonstration that the information content of raw measurements, distributed over multiple samples, are encoded into its internal feature while it also provides a denoising effect; 3) Empirical demonstration that not only does SCARL not require prior knowledge on the leakage model, but that it is also significantly more efficient than model-based attacks on the Ascon authenticated cipher; and 4) Demonstration that the power consumption during S-box computations in the Initialization Stage of Ascon leaks secret information that can be used to recover the entire secret key.

The rest of the paper is organized as follows. A brief introduction to the mathematical background of SCA techniques and deep learning SCA is presented in Section \ref{sec:background}. In Section \ref{sec:info_auto}, we introduce the LSTM autoencoder and show that it extracts information content of all power samples. The SCARL attack with a description of the RL algorithm is provided in Section \ref{sec:scarl}. The practical implementation of attack on Ascon is explained in Section \ref{sec:ascon}, and the experimental results are shown in Section \ref{sec:results}. The paper concludes in Section \ref{sec:conclude}.

\section{Background and Related Work} \label{sec:background}
\subsection{Attack Model} \label{sec:clusterdiff}
In a typical power side-channel analysis, the power consumption corresponding to a key-dependent operation is used to recover the secret key. Assume the cipher operation $F_k(): \mathbb{F}_2^n \to \mathbb{F}_2^m$ maps known input data $Z \in \mathbb{F}_2^n$ to an unknown intermediate variable $X \in \mathbb{F}_2^m$, called the sensitive or secret data, in which $k$ is a subset of the secret key. The function $F_k()$ usually involves a non-linear, or \textit{confusion}, operation in a cryptographic algorithm for a successful key recovery. In most block ciphers, the confusion operation is the so-called substitution layer, or S-box. One class of PA techniques uses the power consumption during the computation of $F_k()$ to find the secret data $X$. Other techniques inspect the power consumption when the variable $X$ is loaded into a memory element in the hardware.

In a cryptographic algorithm, the mutual information between the secret data and the known input/output of the operation under attack, i.e. $F_k()$, is zero when the secret key is unknown. Formally, $H(\Bar{\mathrm{\mathbf{x}}}_r|Z) = H(\Bar{\mathrm{\mathbf{x}}}_r)$ in which $H()$ is the Shannon entropy and $\Bar{\mathrm{\mathbf{x}}}_r$ is an arbitrary combination of $r\in [1,m]$ bits in the binary representation $\Bar{\mathrm{\mathbf{x}}}=(x_i)_{i=0,1,\cdots,m}$ of the secret data $X$. When the key $k$ is unknown, i.e. uniformly distributed with maximum entropy, the distribution of $\Bar{\mathrm{\mathbf{x}}}_r$ is also uniform for a given $Z$. Hence, the mutual information between $Z$ and $X$ is zero. In power analysis, it is assumed that the mutual information between a measured power trace $\mathrm{\mathbf{T}}\in \mathbb{R}^N$ and the secret data $\Bar{\mathrm{\mathbf{x}}}$ is nonzero; i.e., $I(\mathrm{\mathbf{T}}; \Bar{\mathrm{\mathbf{x}}}) > 0$, in which $I(a;b)$ is the mutual information between random variables $a$ and $b$.

The above properties constitute the core of power analysis techniques. A set of input data $Z$ with the corresponding power traces $\mathrm{\mathbf{T}}$, measured during the computation of $X=F_k(Z)$, is available. For every possible value of the key $k$, the values of the secret data $X$ corresponding to a set of known input $Z$ are calculated. Let $\Bar{\mathrm{\mathbf{x}}}_{k^*}$ denote the binary representation of the output of the operation $F_k()$ with a key candidate $k^*$. If $k^*$ is the correct key, we have $I(\mathrm{\mathbf{T}}; \Bar{\mathrm{\mathbf{x}}}_{k^*}) > 0$ which is the basic assumption in side-channel analysis. However, with an incorrect value of the key, $I(\mathrm{\mathbf{T}}; \Bar{\mathrm{\mathbf{x}}}_{k^*}) = 0$. This can be inferred from the fact that $\Bar{\mathrm{\mathbf{x}}}_{k^*}$ is uniformly distributed for different values of $k^*$. If we take the mutual information for ranking the key candidates, the correct key exhibits the highest rank.

The binary representation of the secret data is not convenient for power analysis, since, the power consumption is not necessarily correlated with the individual bits of the data. The numerical normal form \cite{whitnall2014myth} is a generic representation of a Boolean variable that captures the bit dependencies of the data. In this form, the $m$-bit Boolean variable $\Bar{\mathrm{\mathbf{x}}}=(x_i)_{i=0,1,\cdots,m}$ is represented by the monomials $X^U=\prod_{i=1}^m x_i^{u_i}$ of degree $d=HW(U)$, in which, $U \in \mathbb{F}_2^m\backslash\{\mathbf{0}\}$ and $HW(U)$ is the Hamming weight of $U$. To define the mutual information between the power traces and the secret data, the leakage function $L(): \mathbb{F}_2^m \to \mathbb{R}$ is introduced to map the Boolean space of the data to the real-valued space of power measurements. A generic leakage model thus can be defined as 
\begin{equation} \label{eq:leakmodel}
    L(X) = \alpha_0 + \sum_{U\in \mathbb{F}_2^m\backslash \{0\}} \alpha_U X^U + \epsilon 
\end{equation}
in which $\alpha_U\in \mathbb{R}, U\in\mathbb{F}_2^m$ are real-valued coefficients of the model, and $\epsilon$ is a noise term. 

Model-based power analysis techniques are based on finding the maximum correlation between the data leakage of~(\ref{eq:leakmodel}) and the power measurements. The mutual information between $L(X)$ and power traces $\mathrm{\mathbf{T}}$, i.e. $I(\mathrm{\mathbf{T}}; L(X))$, is used as the rank statistics in MIA. To calculate the mutual information empirically, the histogram of the measurements and the leakage can be used to estimate the joint and marginal distributions. Alternatively, in a clustering-based MIA, the power traces are clustered according to the leakage model. The difference between the Shannon entropy of the power traces and the average conditional entropy of traces in the clusters provides the mutual information between measurements and data \cite{whitnall2011exploration}.

The mutual information rank statistics involve a logarithm function that is sensitive to noise and estimation errors. Rather than entropy of measurements, Kolmogorov–Smirnov (KS) statistics calculate the difference between the cumulative distribution function (CDF) of the measurements and their conditional CDF given data. Assume $\mathcal{C}$ denotes a set of clusters of measurements corresponding to particular values of data, the KS statistics for ranking the key candidates $k^*$ is
\begin{equation} \label{eq:KSrank}
    R_{KS}(k^*) = \mathrm{E}_{c\in\mathcal{C}} \Big [ \sup_{t} \big | F_{\mathrm{\mathbf{T}}}(t) - F_{\mathrm{\mathbf{T}}}(t|L(X)\in c) \big| \Big]
\end{equation}
in which $\mathrm{E}_{\mathcal{C}}$ is the expectation over $\mathcal{C}$ and $F_{\mathrm{\mathbf{T}}}(.)$ is the CDF of power traces $\mathrm{\mathbf{T}}$. The KS rank as defined in (\ref{eq:KSrank}) can be considered as a measure for the mutual information between measurements and data. As shown in \cite{whitnall2011exploration}, the KS rank is more robust than the empirical mutual information in the presence of measurement noise.

According to the KS statistics, the correct key exhibits the highest difference between the measurement distribution, i.e. $F_{\mathrm{\mathbf{T}}}(t)$, and the cluster distribution (conditional distribution given a cluster), i.e. $F_{\mathrm{\mathbf{T}}}(t|L(X)\in c)$. We know that
\begin{equation}
    F_{\mathrm{\mathbf{T}}}(t) = \sum_{c\in\mathcal{C}} F_{\mathrm{\mathbf{T}}}(t|L(X)\in c) P\{L(X)\in c\}
\end{equation}
Hence, we can consider the CDF of the measurements as the mean of cluster distributions; thus, the KS statistics are the average difference between the cluster distributions and the mean. Further, since $F_{\mathrm{\mathbf{T}}}(t)$ is the mean distribution, we have
\begin{equation}
    \forall t, \exists\; c_1, c_2 \in\mathcal{C}: F_{\mathrm{\mathbf{T}}}(t|L(X)\in c_1) < F_{\mathrm{\mathbf{T}}}(t), F_{\mathrm{\mathbf{T}}}(t|L(X)\in c_2) > F_{\mathrm{\mathbf{T}}}(t)
\end{equation}
As a conclusion, the maximum difference between the cluster distributions and the mean implies maximum inter-cluster difference. The inter-cluster difference is the most popular rank statistic in the literature of clustering model-based techniques. 
According to the above discussion, the inter-cluster difference is a measure for the information content of the power measurements about the secret data. However, rather than the CDF function, it is more convenient to use the statistical moments of the measurements in practice. In first-order attacks, the difference between the first moments, i.e. the mean, of power measurements in the clusters is used to rank the key candidates. Higher order moments has also been used as a measure for inter-cluster difference \cite{prouff2009statistical, standaert2008partition}.

\subsection{Dimensionality Reduction}
The power traces $\mathrm{\mathbf{T}}=\{t_i|i=0,1,\cdots, N-1 \}$ are $N$-dimensional real-valued vectors corresponding to $N$ samples of the power consumption during computation of the secret variable $X$. However, the domain of data leakage in (\ref{eq:leakmodel}) is the one-dimensional real space. Classical model-based techniques, such as DPA and CPA, assume that one sample of the power traces conveys information about the secret data. In other words,
\begin{equation} \label{eq:poi}
    \exists\; i: 0\le i \le N-1, I(t_i; L(X)) > 0
\end{equation}
This is the main assumption in sample-based techniques; they assume the information about the secret data is concentrated on one sample of the power traces, the so-called point of interest (POI). As a result, if the information of the secret data is distributed over multiple samples, these techniques require an excessively large number of measurements to identify the correct data, since only part of the information is observed. Further, the measurement noise might constructively add to a large inter-cluster difference of a power sample different from the POI when an incorrect key candidate is used for clustering.

One solution to extract information content of the measurements distributed over multiple dimensions (samples of power traces) is to use a dimensionality reduction algorithm. Principal component analysis (PCA) is widely used in SCA techniques based on classical machine learning algorithms for this purpose \cite{lerman2011side}. Assume a number of $S$ power traces $\mathrm{\mathbf{T}}_j, j=1,2,\cdots,S$ are available. The eigenvector decomposition of the sample covariance matrix of the measurements is obtained as $C_{TT}= Q\Lambda Q^t$, in which, $\Lambda$ is a diagonal matrix containing the eigenvalues of the covariance matrix and the columns of matrix $Q$ are the orthonormal eigenvectors. Let the eigenvalues be denoted by $\lambda_0\ge\lambda_1\ge\lambda_2\ge\cdots\ge\lambda_{N-1}$ with the corresponding eigenvectors $\mathbf{q}_i, i=0,1,\cdots,N-1$. The eigenvalue $\lambda_i$ represents the variance, or \textit{power}, of the measurements in the direction of $\mathbf{q}_i$.

The set of eigenvectors of the covarince matrix in PCA constitute an orthonormal basis that spans the space of measurements. The eigenvector corresponding to the largest eigenvalue is the direction with the largest information content in the measurements. Assuming the variance of measurement noise is $\sigma_n^2$, the $i$-th eigenvalue is $\lambda_i=\sigma_{T,i}^2+\sigma_n^2$, in which $\sigma_{T,i}^2$ is the power (variance) of the signal in the direction of $\mathbf{q}_i$. As a result, the signal to noise ratio (SNR) in the $i$-th direction can be obtained as $\lambda_i/\sigma_n^2=1+SNR_i$. This implies that measurement noise is dominant in the directions with smaller eigenvalues resulting in larger estimation errors using the components of the measurements in these directions. 

In PCA, a power trace $\mathrm{\mathbf{T}}$ is projected onto the direction of the eigenvectors with larger eigenvalues which constitute the principal components of the measurements. The power traces are, hence, represented by the components $T_i=\mathrm{\mathbf{T}}^t \mathbf{q}_i, i<N-1$. Instead of individual samples of the power traces, as in relation (\ref{eq:poi}), we now assume that at least one of the principal components $T_i$ has large mutual information with the secret data. We point out that every principal component $T_i$ is now a (linear) function of all samples of the power traces. Further, since the measurements are projected onto directions with larger SNR, the effect of noise is reduced.

While PCA captures distributed information and reduces the effect of noise, it has major limitations undesirable for SCA. First, PCA requires calculating the sample covariance matrix. This implies that the measurements are precisely aligned, hence, the attacker must have precise control over the timing of the hardware under attack. Second, PCA is a linear mapping from the measurements to a reduced dimension space. Therefore, it limits the space of all possible mappings to a small space of linear functions. We employ an autoencoder, based on deep neural networks, as a dimensionality reduction function with the following advantages; 1) Deep neural networks can model a large set of nonlinear functions thus providing a more generic mapping to a reduced-dimension space; 2) We demonstrate that the autoencoder extracts the maximum information content of the measurements while it also reduces the noise; 3) The autoencoder does not require precise alignment of measurements.

\subsection{Deep Learning SCA (DL-SCA)}
\begin{figure}
	\centering
	\includegraphics[width=1\textwidth]{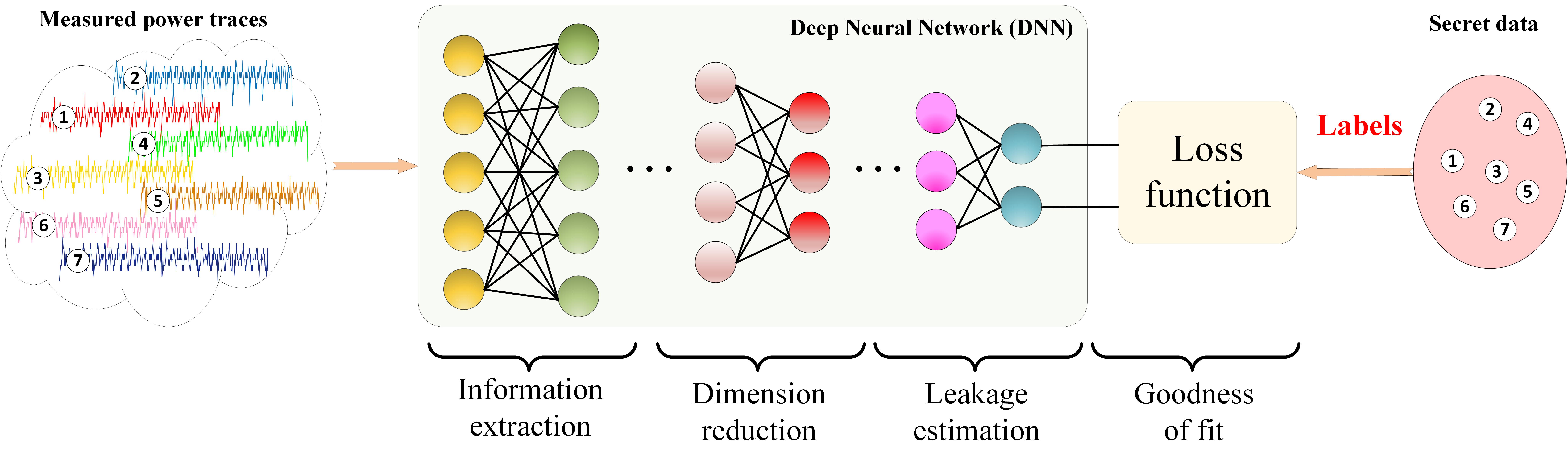}
	\vspace{-0.6cm}
	\caption{Abstract representation of a typical deep learning SCA (DL-SCA) technique for mapping information of power measurements to secret data.}
	\label{fig:dlsca}
\end{figure}
In supervised DL-SCA techniques, all stages of power analysis, i.e. dimensionality reduction, leakage estimation and rank statistics, are implemented with a single DNN. An abstract representation of DL-SCA techniques is shown in Fig.~\ref{fig:dlsca}. A typical DNN consists of multiple hidden layers that successively extract information from the measurements. In the first layers of the DNN, the information of measurements, distributed over multiple samples, is extracted. The intermediate layers provide a compact representation of the information, equivalent to dimensionality reduction in classical machine learning. The final layers map the extracted information to the space of secret data, equivalent to leakage estimation. The cost function of the DNN serves as a measure for the goodness of fit. 

A supervised DL-SCA requires a training set consisting of power measurements with the corresponding secret data. The important point in training the DNN is choosing the proper representation of the Boolean secret data that serves as the labels in the training process. The most popular label is the \textit{one-hot} encoding in which an $m$-bit binary data is represented as a vector of size $M=2^m$ denoted by $\mathbf{l}=(l_0,l_1,\cdots,l_M)$. For a secret variable $X=j\in\mathbb{F}_2^m$, we have $l_j=1$ and $l_i=0, i\ne j$. This encoding is widely used in DL-SCA \cite{maghrebideep, wangtandem, wang2019diversity}. The DNN is trained to generate outputs which are as close as possible to the labels.

The one-hot encoding of data is popular in multi-class learning problems where the last (output) layer of the DNN involves a \textit{softmax} activation function. The output of the softmax function is a vector of $M$ real values in $[0, 1]$ such that the sum of the values is equal to 1. As such, the output of the softmax function can be considered as a probability mass function. In this interpretation, the $j$-th output of softmax is equivalent to the probability that the input powers corresponds to the data value $X=j$. The cross-entropy between the labels and the output of the softmax is used as the loss function for training the DNN. The cross-entropy loss also serves as the goodness of fit statistic during inference.

The labels used for representing the secret data in DL-SCA has significant effect on the leakage model estimated by the DNN. Using one-hot encoding implies that the power consumption corresponding to different values of the secret data are distinct. As a result, the DNN extracts those features of the power traces that result in distinct leakage values for every value of the secret data. This increases the risk of \textit{overfitting} which is a common issue in deep learning~\cite{srivastava2014dropout}. As an example, if the power consumption is correlated with the Hw of data, the power traces measured during the computation of $X_1=0\mathrm{x}AA$ and $X_2=0\mathrm{x}55$ are similar. However, with one-hot encoding, the DNN should learn different features from the power traces to assign $X_1$ and $X_2$ to different classes. In a study in \cite{wang2019diversity}, an MLP neural network is trained with power traces from an ATxmega128D4 microcontroller to recover the secret key of the Advanced Encryption Standard (AES). When the MLP is used to infer the secret data with power traces collected from the same microcontroller mounted on a different board, the accuracy of prediction drops from 88.5\% to less than 13.7\%. This implies overfitting in learning the proper leakage model.

An alternative to one-hot encoding of the secret data is using the leakage function of (\ref{eq:leakmodel}) as the labels, assuming that the leakage model is known a priori. This results in the class of model-based DL-SCA, an example of which is introduced in~\cite{timon2019non}. The model-based DL-SCA can recover secret data without requiring a training set, since the information about the data leakage, that should be estimated from the training set in a profiling DL-SCA, is already incorporated in the leakage model. Hence, the non-profiled DL-SCA of \cite{timon2019non} does not require a training set to recover the secret data, however, prior knowledge of the leakage model is necessary for a successful attack.

In this work, we introduce a DL-SCA technique, based on unsupervised learning, that requires neither a training set nor a full knowledge of the leakage model. We employ an LSTM autoencoder to extract features from the power measurements. We cluster the power features with reinforcement learning (RL) using actor-critic networks. In the following, a brief overview on the basics of RL and the required objective functions to train the actor and critic networks is presented. In the next section, the architecture of the autoencoder that extracts the maximum information of  the measurements is introduced.

\subsection{Reinforcement Learning (RL)} \label{sec:RL}
Deep reinforcement learning (RL) is a class of self-supervised learning techniques to find an optimal policy of control on an environment with high dimensional sensory inputs. RL has been using in algorithms that learn computer games, such as Atari and Go, which exceed the performance of a human \cite{mnih2015human, silver2016mastering, silver2018general}. It has also been used in robotic navigation and autonomous driving to learn optimal policies \cite{kretzschmar2016socially, kahn2018self}.

The main concepts in the terminology of RL include the \textit{actor policy}, the \textit{environment} and the \textit{reward}. The actor takes actions, according to a policy, that changes the state of the environment. The policy is a learning algorithm, such as neural networks. A reward is associated to an action for every state transition in the environment. The state of the environment, and the corresponding reward are observed by the actor which is, in general, a high dimensional input. The goal of RL problem is to find the optimal policy such that after a succession of actions, the actor receives the maximum expected reward.

Let $r(s_i, a_i)$ denote the reward an actor receives by taking action $a_i$ in the state $s_i$. The goal of RL is to obtain the maximum reward after taking a total of $T$ actions. Hence, the instantaneous rewards associated with individual actions is not a suitable measure for training the policy. Rather, we define the discounted reward, associated with the state $s_t$, as $R_t=\sum_{i=t}^T\gamma^{i-t}r(s_i,a_i)$, in which $\gamma$ is the discount factor. The discounted reward is an average of all future rewards obtained after time step $t$. The optimization goal is now to maximize the mean of the discounted reward associated with the initial state, i.e. $J=E[R_1]$, in which the expectation is over the space of all actions, rewards and states of the environment. 

Most classical RL algorithms use so-called Q-learning to optimize the policy. The Q-function, or the action-value function, is defined in terms of the discounted reward as
\begin{equation}
    Q^{\pi}(s_t, a_t) = \mathrm{E}_{r_{i\ge t}, s_{i>t}, a_{i>t}}[R_t|s_t,a_t]
\end{equation}
In this relation $\pi:\mathcal{S} \to P(\mathcal{A})$ is the policy function that maps the space of states to the probability distribution over the space of actions. The Q-value, associated with a policy $\pi$, is the conditional expectation of the discounted reward associated with action $a_t$ in state $s_t$ at step $t$, when the actions are sampled from the output of the policy $\pi$. The expectation is over the space of future actions and states after the step $t$.

When the space of states and actions is finite, the Q-values of the optimal policy can be calculated using an iterative procedure called the Bellman equation \cite{lillicrap2015continuous}. However, in the application of clustering the measurements for SCA, the size of state space is practically infinite, as explained later. However, Q-learning requires the discounted rewards of all possible states. A common technique to address this issue is using a deep neural network to estimate the discounted reward associated with an observed state. These are called deep Q-network (DQN) techniques.

The Q-value is the mean discounted reward given that action $a_t$ is taken at state $s_t$. The expectation of the Q-values over all actions in the state $s_t$ is defined as the state-value, which is expressed as $V^{\pi}(s_t)=\mathrm{E}_{a_t\sim\pi}[Q^{\pi}(s_t, a_t) ]$.
The advantage function is then defined as 
\begin{equation}
    A^{\pi}(s_t, a_t) = Q^{\pi}(s_t, a_t) - V^{\pi}(s_t)
\end{equation}
The advantage of action $a_t$ in state $s_t$ is the mean discounted reward of the action in the state minus the expectation of the discounted reward over all actions in the state. Similar to Q-values in DQN, the state-value of an observed state can also be estimated using neural networks. Assume the parameters of the neural network are $\theta^V$. The loss function for finding the optimal parameters is
\begin{equation} \label{eq:tdmse}
    \mathcal{L}(\theta^V) = \mathrm{E}_{r_t, s_t\sim\rho^{\pi}} \big[\big(y_t - V(s_t|\theta^V)\big)^2 \big]
\end{equation}
in which $\rho^{\pi}$ is the \textit{visitation} probability of an state with the policy $\pi$. Further,
\begin{equation} \label{eq:target}
    y_t = r(s_t, a_t) + \gamma\cdot V(s_{t+1}|\theta^V)
\end{equation}
is a measure for the state-value. We note that $y_t$ also depends on the parameters $\theta^V$ which are usually ignored in the optimization problem. The difference $\delta_t=y_t - V(s_t|\theta^V)$ is called \textit{time difference} (TD) error. The learning algorithms based on minimizing the MSE of the TD error, as in (\ref{eq:tdmse}), are called TD-learning.

The neural network used as the state-value estimator is called the \textit{critic} network, for the reasons explained later. Next, we need the proper objective for finding the optimal policy. Let the parameters of the policy neural network be $\alpha$. Policy gradient algorithms update the parameters in the direction following the gradient of the mean reward. The gradient of the mean reward is \cite{heess2012actor}
\begin{equation} \label{eq:gradpolicy}
    \nabla_{\alpha}J(\alpha) = \mathrm{E}_{s_t\sim\rho^{\pi}, a_t\sim\pi}\big[ A(s_t, a_t|\alpha) \nabla_{\alpha} \log\pi(s_t,a_t|\alpha) \big]
\end{equation}
Further, it can be shown that the TD error $\delta_t$ is an unbiased estimate for the advantage function $A(s_t, a_t|\alpha)$ \cite{heess2012actor}. Replacing the advantage function in (\ref{eq:gradpolicy}) with $\delta_t$ provides the policy gradient for training the policy network. According to this relation, the policy network (actor) wishes to follow the direction of the gradient. However, $\delta_t$, as calculated by the critic network, determines whether this direction achieves a higher discounted reward ($\delta_t>0$), or the opposite direction should be taken ($\delta_t<0$). This is why the state-value estimator is called the critic network as it provides a critique for the actions of the policy network.

\section{Maximum Information Extraction} \label{sec:info_auto}
According to the discussion of the previous section, a necessary step in a successful SCA attack is to encode the information content of the raw power measurements into an intermediate representation. The major reason is to capture all information distributed over multiple dimensions, which possibly reduces the effects of noise. This step is inherent in training the DNNs used in supervised or model-based DL-SCA; the internal neural features of the DNN include all relevant information of the input. The autoencoder is a powerful unsupervised learning technique for encoding an input data into an intermediate representation.

\subsection{Autoencoder} \label{sec:autoecnoder}
An autoencoder learns an intermediate representation of the input data by mapping the data into an internal feature space and reconstructing the original data from the features. If the reconstruction error is small, the internal representation includes all information about the original data. Let assume $\hat{\mathrm{\mathbf{T}}}$ is the input to the autoencoder which is a version of an original data $\mathrm{\mathbf{T}}$ corrupted by a random process. The autoencoder consists of an encoder $e():\mathbb{R}^N\to \mathbb{R}^D$ which maps the input to a $D$-dimensional feature space, and a decoder $d():\mathbb{R}^D\to \mathbb{R}^N$ which maps the feature space back into the space of original data. The encoder and decoder are parameterized by $W_e$ and $W_d$. The optimal parameters are obtained by minimizing a loss function $\mathrm{E[\mathcal{L}(\mathrm{\mathbf{T}}, \hat{\mathrm{\mathbf{T}}})]}$.

Depending on the optimization objective, the autoencoder can be considered as different learning algorithms. 
The general form of the optimization problem for a regularized autoencoder can be expressed as 
\begin{equation} \label{eq:cae}
    e,d = \argmin_{e,d} \mathrm{E}\Big[\mathcal{L}(\mathrm{\mathbf{T}}, \hat{\mathrm{\mathbf{T}}}) + \lambda \mathcal{R}(\mathrm{\mathbf{T}}, \hat{\mathrm{\mathbf{T}}}) \Big]
\end{equation}
in which $\mathrm{E[\mathcal{R}(\mathrm{\mathbf{T}}, \hat{\mathrm{\mathbf{T}}})]}$ is a regularization term. If the regularization term is $\big\| \partial e(\hat{\mathrm{\mathbf{T}}})/\partial \hat{\mathrm{\mathbf{T}}} \big\|^2_F$, as in contractive autoencoder (CAE) \cite{rifai2011contractive}, it learns data manifold. Due to the regularization term, the sensitivity of the encoder function on the manifold, around which the measurements are distributed, is minimized. As a result, the autoencoder learns the normal direction to the data manifold. The denoising autoencoder (DAE) \cite{vincent2008extracting} does not include an explicit regularization term. The autoencoder is optimized with a training set consisting of the original data and the corresponding version corrupted with a random process. The function $\mathcal{L}$ is the cross-entropy loss for binary data. The training set regularizes the autoencoder, implicitly. Further, \cite{alain2014regularized} has introduced the reconstruction contractive auto-encoder (RCAE) with the regularization on the encoder and decoder function composition, i.e. $r()=d\circ e()$. It is shown a Gaussian process, the RCAE learns the score function from which the input distribution can be estimated. In the following, we demonstrate that using the maximum likelihood criterion as the optimization goal, and minimizing the entropy as the regularization, the autoencoder extracts the maximum information from the input.

The goal of a maximum information autoencoder is to extract features $\mathbf{f}\in\mathbb{R}^D$ from the corrupted measurements $\hat{\mathrm{\mathbf{T}}}$ which have the highest mutual information with the original data $\mathrm{\mathbf{T}}$. Formally, we want to find the optimal parameters $W_e$ of an encoder function $e()$ to maximize
\begin{equation} \label{eq:mutualinfo}
    I(\mathrm{\mathbf{T}};\mathrm{\mathbf{f}}) = H(\mathrm{\mathbf{T}}) - H(\mathrm{\mathbf{T}}|\mathrm{\mathbf{f}})
\end{equation}
In the above equation, $H(\mathrm{\mathbf{T}})$ is the Shannon entropy of the original data which is independent of the parameters $W_e$. Hence, the optimal encoder parameters can be obtained by minimizing the conditional entropy. 
Using the definition of the conditional entropy into the above equation, and considering that the output of the decoder is an estimate of the original data, we have
\begin{equation} \label{eq:minentropy}
    \hat{W}_e = \argmax_{W_e,\Tilde{p}} \mathrm{E}_{(\mathrm{\mathbf{T}},\mathrm{\mathbf{f}})}[ \log\Tilde{p}(\mathrm{\mathbf{T}}|\mathrm{\mathbf{f}}) ]
\end{equation}
We note that the conditional probability function $\Tilde{p}(\mathrm{\mathbf{T}}|\mathrm{\mathbf{f}})$ is restricted by the structure of the decoder. Hence, to find the optimal parameters of the encoder, the optimization problem of (\ref{eq:minentropy}) must be solved over all possible conditional probability density functions $\Tilde{p}$. With the restricted decoder structure, the criterion in (\ref{eq:minentropy}) is a lower bound on the conditional entropy of (\ref{eq:mutualinfo}). 

An alternative perspective to the optimization problem of the maximum information autoencoder is obtained by replacing the the conditional distribution $\Tilde{p}(\mathrm{\mathbf{T}}|\mathrm{\mathbf{f}})$ in (\ref{eq:minentropy}) with $\hat{p}(\hat{\mathrm{\mathbf{T}}}|\mathrm{\mathbf{T}}; W_e,W_d)\hat{p}(\mathrm{\mathbf{T}}; W_e,W_d)$. In this relation, The variable $\mathbf{f}$ is replaced by the measurements $\hat{\mathrm{\mathbf{T}}}$ as $\mathbf{f}=e(\hat{\mathrm{\mathbf{T}}})$ is a deterministic function given the parameters of the encoder. Further, the parameters $W_e$ and $W_d$ are explicitly shown to emphasize that the distributions are restricted by the encoder and decoder structures. Using this replacement, the optimum parameters of the autoencoder are obtained as
\begin{equation}\label{eq:WeWd}
    \hat{W}_e, \hat{W}_d = \argmax_{W_e,W_d} \mathrm{E}_{(\mathrm{\mathbf{T}},\hat{\mathrm{\mathbf{T}}})} \big[ \log\hat{p}(\hat{\mathrm{\mathbf{T}}}|\mathrm{\mathbf{T}}; W_e,W_d)
     +\log \hat{p}(\mathrm{\mathbf{T}}; W_e,W_d) \big]
    = \argmax_{W_e,W_d} \Big\{ \mathrm{E}_{(\mathrm{\mathbf{T}},\hat{\mathrm{\mathbf{T}}})} \big[ \log\hat{p}(\hat{\mathrm{\mathbf{T}}}|\mathrm{\mathbf{T}}; W_e,W_d) \big] - H(\mathrm{\mathbf{T}}) \Big\}
\end{equation}
The first term in this optimization goal is the conditional likelihood of the measurements given the original data. The second term is the Shannon entropy of the original data as reconstructed by the decoder. Comparing (\ref{eq:WeWd}) with (\ref{eq:cae}), the maximum information autoencoder can be considered as a regularized maximum likelihood problem in which the regularization is minimizing the entropy of reconstruction.

The maximum likelihood criterion reduces to mean squared error with Gaussian distribution. Considering that the corruption process in the measurements is the additive Gaussian noise, we have $\hat{\mathrm{\mathbf{T}}}=\mathrm{\mathbf{T}}+\mathrm{\mathbf{N}}$, in which $\mathrm{\mathbf{N}}\sim\mathcal{N}(0,\Sigma)$ and $\Sigma$ is the covariance of the noise. The general form of a multivariate Gaussian distribution with $N$ variables with the mean vector $\mathbf{\mu}$ and covariance matrix $C$ is 
\begin{equation} \label{eq:Gauss}
    f(\mathbf{Y};\mathbf{\mu},C) = \frac{1}{\sqrt{(2\pi)^N|C|}} e^{-(\mathbf{Y}-\mathbf{\mu})^t C^{-1}(\mathbf{Y}-\mathbf{\mu})}
\end{equation}
in which $|C|$ is the determinant of $C$. With this definition, the conditional $\hat{\mathrm{\mathbf{T}}}|\mathrm{\mathbf{T}}$ is $p(\hat{\mathrm{\mathbf{T}}}|\mathrm{\mathbf{T}})=f(\hat{\mathrm{\mathbf{T}}}; \mathrm{\mathbf{T}}, \Sigma)$.
By using this distribution in the regularized maximum likelihood problem of (\ref{eq:WeWd}), we obtain
\begin{equation} \label{eq:MSEloss}
    \hat{W}_e, \hat{W}_d = \argmin_{W_e,W_d}  \Big\{ \mathrm{E}_{(\mathrm{\mathbf{T}},\hat{\mathrm{\mathbf{T}}})}  \big[ (\hat{\mathrm{\mathbf{T}}}-\mathrm{\mathbf{T}})^{T}\Sigma^{-1}(\hat{\mathrm{\mathbf{T}}}-\mathrm{\mathbf{T}}) \big]
    +H(\mathrm{\mathbf{T}}) \Big\} = 
    \argmin_{W_e,W_d}  \Big\{ \mathrm{E} \big[ \|\hat{\mathrm{\mathbf{T}}}-\mathrm{\mathbf{T}}\|^2 \big]
    +H(\mathrm{\mathbf{T}}) \Big\}
\end{equation}
The second equality holds for stationary measurement noise. As a result, for Gaussian noise, the optimization problem of a max information autoencoder reduces to minimizing both the mean squared error (MSE) between the input and output of the autoencoder and the entropy of the output.

The Shannon entropy is a logarithmic function of the distributions which is sensitive to estimation errors. Hence, the entropy term in the optimization objective of (\ref{eq:MSEloss}) makes the training process of the autoencoder difficult. In the following, we demonstrate that a constraint on the dimension of the feature space $\mathbf{f}$, i.e. $D$, set a constraint on the entropy. As a result, minimizing the entropy can be achieved by \textit{hyperparameter} tuning in the autoencoder, in which the hyperparameter is the dimension of the feature space. We show that the constraint on the feature dimension results in a limit on the number of degrees of freedom in the reconstruction, thus, the entropy.

\subsection{Number of Degrees of Freedom (NDF)} \label{sec:ndf}
We demonstrate that the dimension of autoencoder features limit the NDF of reconstructed power traces, which in turn reduces the entropy and filters high frequency components. In estimation theory, the NDF refers to the number of independent variables that represent a random process, or a set of measurements. The NDF is also closely related to the information content of the measurements. In the following, we also show that NDF is inversely proportional to low frequency components of the measurements.

A convenient definition for the NDF in a random process, useful for empirical purposes, based on the sample moments is provided in \cite{kikkawa1994number}. Let $\mathrm{\mathbf{T}}=\{t_i|i=0,1,\cdots, N-1 \}$ denote the samples of an \textit{ergodic} random process $T[n]$ in an interval of length $N$. The $m$-th moment of the random process in the interval can be calculated as
\begin{equation} \label{eq:samppmoment}
    S_{m}(N) = \sum_{i=0}^N t_i^m /N
\end{equation}
The $m$-th order NDF of the random process in the interval of length $N$ is defined as
\begin{equation} \label{eq:ndf}
    k_{m}(N) = Var\{T^m[n] \}/Var\{S_{m}(N) \}
\end{equation}
in which $Var\{.\}$ is the variance. This definition holds for a random process $T[n]$ that is ergodic regarding the $m$-th moment. As a result, $\lim_{N\to\infty}1/k_{m}(N)=0$ which implies that the $m$-th sample moment converges to the ensemble moment in mean.

Next, we show that the first order NDF, as defined in (\ref{eq:ndf}), is directly proportional to the Fisher information of an ergodic Gaussian process. The Fisher information around the first moment of a random process $\mathbf{T}$ with likelihood function $f(\mathrm{\mathbf{T}}; \mathbf{\mu}_T, C_T)$ is defined as 
\begin{equation} \label{eq:fisher}
    I_1 = -E\big[\frac{\partial^2}{\partial\mathbf{\mu}^2}\log f(\mathrm{\mathbf{T}}; \mathbf{\mu}_T, C_T) \big]
\end{equation}
Using the density function in (\ref{eq:Gauss}), it can be shown that the above equation reduces to the sum of the elements in the inverse covariance matrix. By denoting $C_T^{-1}=[a_{ij}]$, we have $I_1=\sum_i\sum_j a_{ij}$. Using this property, we can express the Fisher information in terms of the frequency spectral components of the random process.

The frequency spectrum of the random process can be obtained as the Fourier transform of the covariance function. Expressing the Fourier transform as the matrix $W$ with elements $w_{k,s}=\exp\{-j2\pi ks/N \}$, the spectrum of the random process in an interval of length $N$, with covariance matrix $C_T$, can be obtained as $\Omega_T=WC_TW^H/N$. In this representation, $\Omega_T$ is a diagonal matrix with the diagonal elements $\omega_i, i=0,1,\cdots,N-1$ equal to the frequency spectral components of the random process. We note that the columns of the Fourier transformation $W$ constitute an orthonormal basis. Accordingly, the spectrum of the inverse covariance matrix is $\Omega_T^{-1}=WC_T^{-1}W^H/N$. As a result, the Fisher information in (\ref{eq:fisher}) for a Gaussian process can be obtained as 
\begin{equation} \label{eq:invomeg0}
    I_1 = \sum_i\sum_j a_{ij} = \sum_{i}\sum_j a_{ij}\cdot w_{0i}w^*_{0j} = N\cdot\omega_0^{-1}
\end{equation}

Since $\omega_0$ is an eigenvalue of the covariance matrix $C_T$, corresponding to the Fourier transform basis, we can show that $\omega_0$ is the sum of elements in $C_T$, due to the the symmetric Toeplitz property of the covariance matrix. Using this property and equation (\ref{eq:invomeg0}), it is shown in \cite{kikkawa1994number} that  $k_1(N)=\sigma_T^2I_1$, with the NDF $k_1(N)$ is defined in (\ref{eq:ndf}). In other words, the first order NDF of the random process is proportional to the Fisher information.

The interpretation of the above discussion in the context of the autoencoder follows. First, we note that the first order sample moment of the random process $T[n]$ in (\ref{eq:samppmoment}) is an estimate for the mean of the process with $N$ samples. Hence, $Var\{S_1(N)\}$ is the variance of the estimation error, or the power of error. On the other hand $Var\{T[n]\}$ is the total power (variance) of the process. Therefore, the NDF of (\ref{eq:ndf}) is related to the signal-to-noise ratio. We also showed that the signal-to-noise ratio is proportional to the Fisher information with the proportionality constant equal to the power of the process.

Additional insight into the NDF is obtained using the frequency spectrum of the covariance. The Fourier transform of the covariance matrix decomposes the random process into components based on the correlation time. The first component, i.e. $\omega_0$, refers to the correlation time of the DC component in the process; Larger $\omega_0$ implies that the samples of the process are correlated over a larger time interval. According to (\ref{eq:invomeg0}) and $k_1(N)=\sigma_T^2I_1$, larger correlation time implies lower NDF. Hence, the autoencoder \textit{smoothes} the measurements.

In an autoencoder, the output is reconstructed from a feature of dimensions $D$. Hence, the NDF of the reconstructed process is limited by $D$; using smaller $D$ results in larger correlation time of the reconstruction. Let $\Tilde{\mathbf{T}}$ denote the output of the autoencoder, which follows a multivariate Gaussian distribution with the mean $\Tilde{\mathbf{\mu}}$ and covariance matrix $\Tilde{C}$. According to the distribution (\ref{eq:Gauss}), the Shannon entropy $H(\Tilde{\mathbf{T}})$ can be found as the mean of $\mathbf{e}^t\Tilde{C}\mathbf{e}$, in which $\mathbf{e}=\Tilde{\mathbf{T}}-\Tilde{\mathbf{\mu}}$ is the variation around the mean. The term $\mathbf{e}=\Tilde{\mathbf{T}}-\Tilde{\mathbf{\mu}}$ can be considered as the sum of the powers of the variation $\mathbf{e}$ over the spectral components $\omega_i, i=0,1,\cdots,N-1$. Lower NDF implies that the higher frequency components $\omega_i, i>0$ are attenuated. Hence, the autoencoder filters out high frequency components, and thus acts as a denoising filter. This results in lower entropy $E[\mathbf{e}^t\Tilde{C}\mathbf{e}]$.

The constraint on the dimension of the features space in the autoencoder provides a trade-off between the \textit{bias} and the \textit{variance} of the estimation. We note that the entropy of reconstruction is due to the measurement noise, but also the information of the measurements about the secret data. Lower dimension results in larger bias, as the DC component of the random process is larger. Hence, part of the information about the data might be missed. However, the higher frequency components, mostly contributed by the measurement noise, are reduced which results in smaller variance. To achieve the minimum MSE in estimating the original data $\mathrm{\mathbf{T}}$, the dimension of the feature space should be adjusted properly such as the total error, due to both bias and variance, is minimized.

\subsection{LSTM Autoenconder} \label{sec:lastmauto}
Recurrent neural networks (RNN) has been used mainly in learning temporal models of signals in a wide range of applications including natural language processing (NLP) \cite{yin2017comparative}, sequence labeling, speech recognition and acoustic modeling \cite{sak2014long}. The long short-term memory (LSTM) neural network is a special type of RNN that can learn short-term as well as long-term temporal model of a signal.

\begin{figure}
	\centering
    \includegraphics[width=1\textwidth]{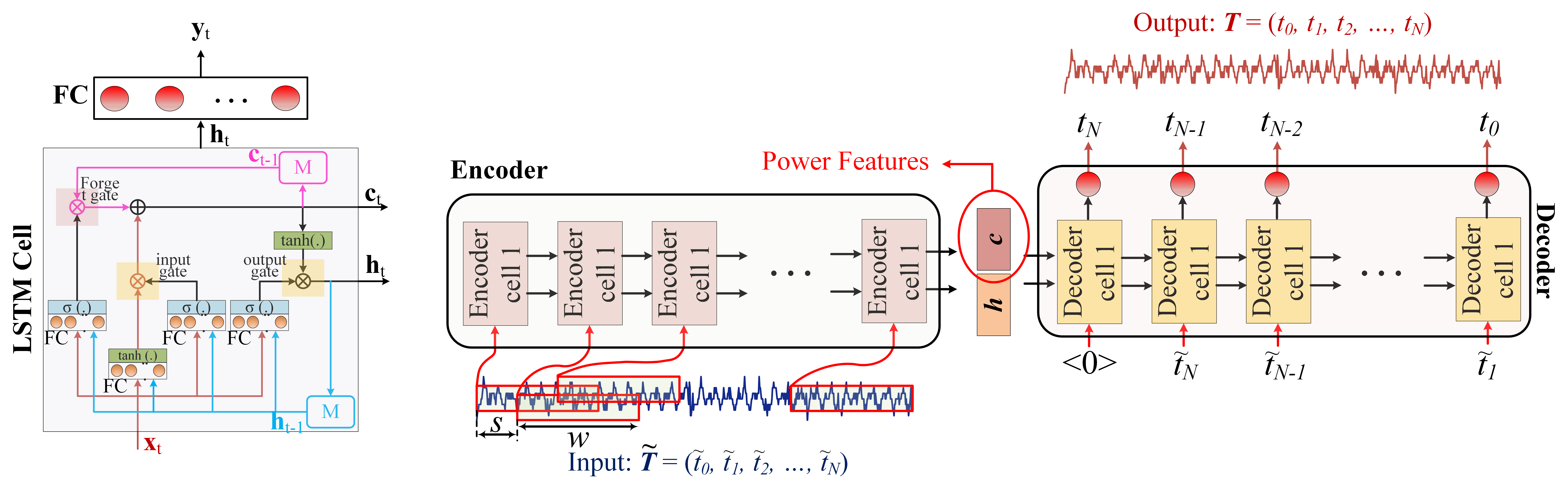}
	\vspace{-0.6cm}
	\caption{Proposed LSTM auto-encoder for extracting features of power traces, with sliding window processing of input power trace and $\mathrm{\mathbf{c}}$ state of top encoder cell selected as power feature.}
	\label{fig:lstm_auto}
\end{figure}

The basic block of an LSTM neural network is shown in Fig. \ref{fig:lstm_auto}. The internal state of the cell encoding information of the input is $\mathrm{\mathbf{c}}$ while $\mathrm{\mathbf{h}}$ is the output at every time instance. The internal state and the output, with equal dimensions, are not bounded by a non-linear activation. In order to generate outputs with different dimensions from the internal state and possibly processed by a non-linear activation, additional neurons are used at the output of the cell. As shown in Fig. \ref{fig:lstm_auto}, a fully connected (FC) layer is added at the output of the cell for this purpose.

A fully connected (FC) layer at the input of the LSTM cell processes the input data. The activation function of the neurons in this layer is $tanh$. The internal state of the cell operate as a memory; a weighted sum of the state at the previous times and the input at the current time update the state. The weight of the previous state and the input are determined by \textit{forget gate} and \textit{input gate}, respectively, which also determine the memory of the cell. The output of the cell, i.e. $\mathrm{\mathbf{h}}_t$, is generated from the internal state, by applying $tanh$ activation first, and then processing the result by \textit{output gate}, which controls the information flow from state to the output. The weights of the three gates, described above, are determined by three FC units with sigmoid activation.

As discussed in Section \ref{sec:autoecnoder}, the structure of the auto-encoder constrains the space of extracted features, hence, a trade-off between variance and bias of estimation. Consequently, the proper structure has a significant role in the success of an unsupervised power analysis. Due to the special properties of LSTM networks in extracting temporal information, we propose an LSTM autoencoder to extract data-dependent features of power traces. The autoencoder is shown in Fig. \ref{fig:lstm_auto}. The encoder and decoder are LSTM networks shown in a \textit{time-unrolled} representation; each of the encoder and decoder consists of a single LSTM cell, however, multiple cells shown in this representation correspond to successive time instants that a cell process the input samples. A single linear neuron is used at the output of the decoder cell to generate samples of a power trace.

The raw power traces are first processed by a sliding window of length $W$ and stride $S$ as shown in Fig. \ref{fig:lstm_auto}. The input vector to the encoder LSTM cell at every time instant is the $W$ samples of successive windows. The process of the sliding window is similar to a convolution layer in CNNs. For faster convergence of the learning process, we also feed the samples of the input power traces to the input of the decoder. The output of the decoder are processed by a single neuron, with linear activation function, to generate the reconstructed power traces, in the reverse order. 

We use the MSE between the intput and output of the autoencoder as the loss function for updating the weights of the encoder/decoder cells. The regularization on the learning process is achieved by tuning the dimension of the internal state as the hyperparameter. The features extracted from the input power traces, are the internal state $\mathrm{\mathbf{c}}$ of the encoder LSTM cell.

\section{SCA with Reinforcement Learning (SCARL)} \label{sec:scarl}
After encoding the information measurements into an intermediate representation, i.e. the autoencoder features, we need a statistic to extract information about the secret data. However, we note that not all all of the features of the autoencoder necessarily relate to the secret data. They include information about all underlying processes of the hardware running in parallel with the cipher operation under attack, including the state machine, clock tree and the peripheral circuitry. The information about the secret data is encoded into at least one of the features.

According to the discussion of Section \ref{sec:clusterdiff}, inter-cluster difference refers to the information content of the measurements about the secret data. Using reinforcement learning (RL) we divide the features of power traces into two clusters such that at least one of the features exhibits the highest inter-cluster difference. Next, we identify the particular feature which corresponds to the secret data. SCARL operates in two steps:
\begin{enumerate}
    \item \textbf{Pronounced Features:} The features of power traces are divided into two clusters satisfying two conditions: 1) the features are evenly distributed over the clusters, and 2) the inter-cluster difference on the mean of at least one feature is maximized. We postulate that at least one of the features with large, but not necessarily the largest, inter-cluster difference corresponds to the secret data.
    \item \textbf{Leakage Estimation:} We label the two clusters as $\{0, 1\}$. We estimate the coefficients of the generic leakage model in (\ref{eq:leakmodel}) such that the leakage values are equal to the labels of the clusters. We postulate that the correct key corresponds to a leakage model in which the low order terms have the largest contribution to the leakage.
\end{enumerate}
\subsection{Actor-Critic Networks} \label{sec:actor-critic}
In this section, we provide a formulation for the problem of clustering the power features in the context of reinforcement learning. 
We use the actor-critic network using the TD advantage learning algorithms, with the objective functions (\ref{eq:tdmse}) and (\ref{eq:gradpolicy}), to assign the power features to two clusters. For this purpose, we provide the following definitions.
\begin{itemize}
    \item \textbf{Actor Policy:} It is a neural network that classifies the power features into two clusters with labels $\{0,1\}$. The action is the class assignment. The output of the neural network is the mean and variance of a Gaussian distribution that defines the stochastic policy. 
    \item \textbf{Environment:} The environment consists of the set of power features grouped into two clusters. The state of the environment is the mean difference between the features in the clusters.
    \item \textbf{Reward:} The reward is composed of two terms. One term is the maximum inter-cluster difference of the mean features. The second term measures the even assignment of features into the clusters.
\end{itemize}

\begin{figure}
	\centering
    \includegraphics[width=0.8\textwidth]{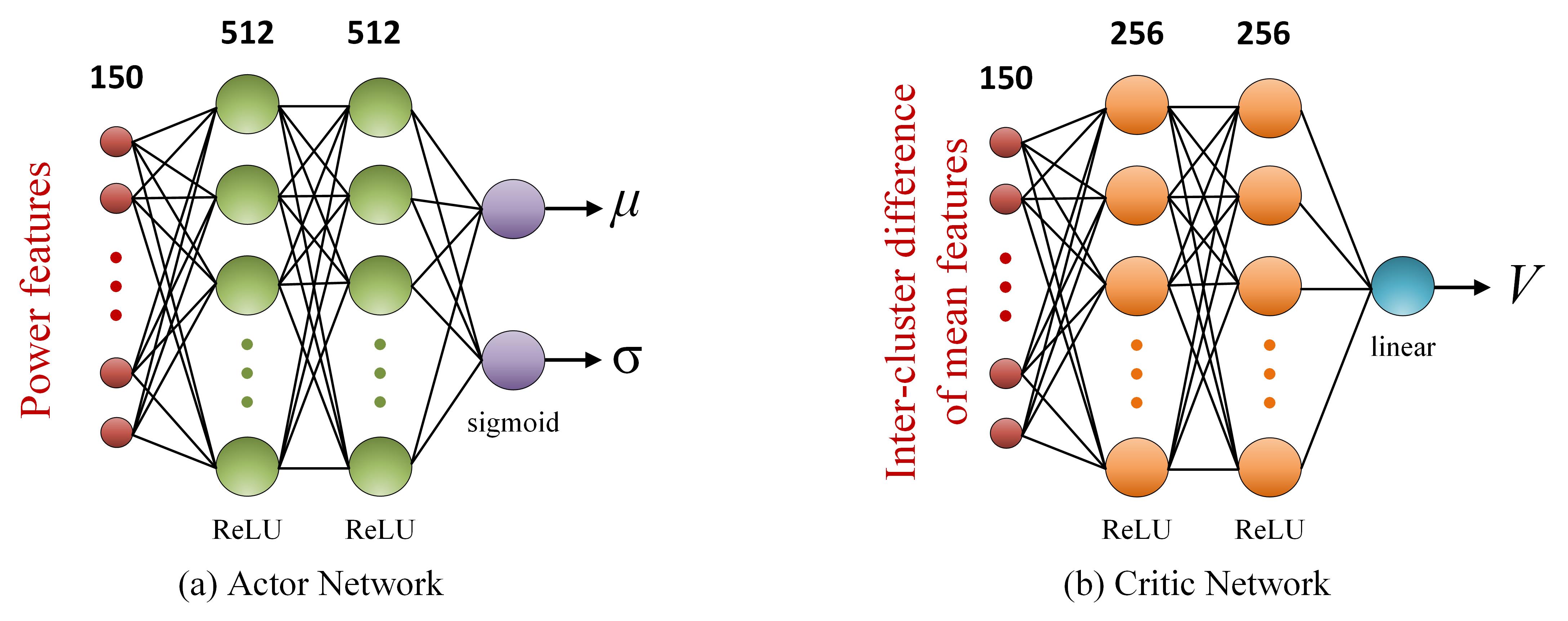}
	\vspace{-0.3cm}
	\caption{Two hidden layer MLP networks used as the actor and critic neural networks for clustering power features with reinforcement learning.}
	\label{fig:actocritic}
\end{figure}
The architecture of the actor and critic networks are shown in Fig. \ref{fig:actocritic}. Both networks are multi-layer perceptrons (MLP) with two hidden layers. Each hidden layer consists of 512 and 256 neurons in the actor and critic networks, respectively. The activation function of the hidden layers is ReLU. The output layer of the actor network consists of two neurons with sigmoid activation. These neurons correspond to the mean $\mu$ and variance $\sigma$ of a Gaussian distribution from which the actions are sampled. The input of the actor network is the features of the power traces, obtained by the autoencoder.

Assume a number of $S$ power traces $\mathbf{T}_j$ with the corresponding features $\mathbf{c}_j,j=1,2,\cdots,S$ are available. The vectors $\mathbf{c}_j$ are the internal state of the LSTM autoencoder described in Section \ref{sec:lastmauto}, which are not bounded by a non-linear activation function. Usually, the training of neural networks are successful if the inputs are normalized. Hence, we calculate the normalized power features as
\begin{equation} \label{eq:pfeatures}
    \Tilde{\mathbf{f}}_j = \frac{\mathbf{c}_j-\mathrm{E}_j[\mathbf{c}_j]}{\max\limits_{j}(\mathbf{c}_j)-\min\limits_{j}(\mathbf{c}_j)}
\end{equation}
in which the expectation, min and max functions are element-wise. The normalized power features $\Tilde{\mathbf{f}}_j$ are used as the input of the actor network. The corresponding outputs of the actor are $\mu_j,\sigma_j,j=1,2,\cdots,S$. The class assignment of every power feature $\Tilde{\mathbf{f}}_j$ is then sampled from the Gaussian distribution with the mean and variance equal to $\mu_i$ and $\sigma_j$, respectively. In other words, $a_j\sim\mathcal{N}(\mu_i,\sigma_j)$ determines the cluster of the $j$-th power measurement. 

Assume that we intend to divide the power features into two clusters $C_0$ and $C_1$. Using the outputs of the actor network, if $a_j<0.5$, we assign $\Tilde{\mathbf{f}}_j\in C_0$ and if $a_j>0.5$ then $\Tilde{\mathbf{f}}_j\in C_1$. Next, we define the state of the environment. Given the actions of the actor network is $a_j$ at time step $t$, the state is the mean difference between the power features in two clusters, i.e.
\begin{equation} \label{eq:meandiff}
    s_t = \mathrm{E}_{\mathbf{c}_j\in C_0}\big[\mathbf{c}_j \big] - \mathrm{E}_{\mathbf{c}_j\in C_1}\big[\mathbf{c}_j \big]
\end{equation}
The state is the input to the critic. Similar to the actor, we need to normalize the input of the critic network. Hence, we use $\Tilde{s}_t=s_t/(\max(s_t)-\min(s_t))$ as the input of the critic.

At every state of the environment, we need to define a reward associated with a action. We define a reward comprised of two terms measuring how well the actor separates the power traces and how evenly distribute them in the clusters. Let $|C_k|,k=0,1$ denote the size of the clusters. We define the percentage of the clusters as $p_k=|C_k|/S, k=0,1$. We recall that $S$ is the total number of measurements. If the power features are evenly distributed, then $p_0=p_1=0.5$. We use the Kullback-Leibler (K-L) divergence between the distribution $P=\{p_0,p_1\}$ and uniform distribution to measure the even assignment of clusters. The reward is then
\begin{equation} \label{eq:reward}
    r(s_t, a_t) = \max (|s_t|) - \mathcal{D}_{KL}(Q||P)
\end{equation}
in which $|s_t|$ is the absolute value of the mean difference between the clusters, as defined in (\ref{eq:meandiff}). The first term of the reward measures the inter-cluster difference. The second term $\mathcal{D}_{KL}(Q||P) = -(1+\mathrm{E}_Q[\log P])$ is the K-L divergence between $P=\{p_0,p_1\}$ and the uniform distribution $Q=\{0.5,0.5\}$.

The actor and critic MLP networks are trained using the objective functions defined in Section \ref{sec:RL}. At every time step $t$, the power features are assigned to either of two clusters $C_0$ and $C_1$ by the stochastic actor policy, as described above. The new state and and the associated reward are calculated using (\ref{eq:meandiff}) and (\ref{eq:reward}), respectively. The target state-value $y_t$ and the corresponding TD error $\delta_t$ are calculated according to (\ref{eq:target}). The critic MLP is updated by minimizing the loss function (\ref{eq:tdmse}). Finally, the parameters of the actor MLP are updated according to (\ref{eq:gradpolicy}). The procedure continues until the obtained reward does not change significantly.

\subsection{Leakage Estimation} \label{sec:loworderleak}
The RL-based clustering algorithm, in the previous section, detects the pronounced features of power measurements, i.e. those features that exhibits the highest inter-cluster difference. At least one of these features corresponds to the secret data. To identify which feature contains information about the secret data, we first estimate a non-linear leakage model for all key candidates. The key candidate that exhibits the highest inter-cluster difference with lower order terms is the correct key.

Assume the value $l_j$ is the action assigned by the actor MLP to the power feature $\mathbf{c}_j$. Hence, $l_j$ refers to the likelihood that the feature belongs to a cluster. Let $Z_j$ denote the (known) input data to the cipher operation under attack associated with the power feature $\mathbf{c}_j$. for every key candidate $k^*$, the secret data $X_j^*$ corresponding to the power feature $\mathbf{c}_j$ is calculated as $X_j^*=F_{k^*}(Z_j)$. This is described in Section \ref{sec:clusterdiff}. Now, the coefficients of the generic leakage model (\ref{eq:leakmodel}) is estimated for the key candidate $k^*$ using MSE; i.e.,
\begin{equation}
    \alpha_U^{*} = \min_{\alpha_U} \mathrm{E}_{j}\big[|L(X_j^*) - l_j|^2 \big], \; U\in\mathbb{F}_2^m
\end{equation}

For an $m$-bit secret data $X$, the highest order of the leakage model is m. We postulate that the components of the leakage model with order $d=HW(U)\le m_0$ with $m_0<m$ have the largest contribution to the leakage. Hence, we set $\alpha_U^{*}=0, HW(U)>m_0$ and find the low order leakage as
\begin{equation} \label{eq:lowmodel}
    l^*_j = \alpha_0^* + \sum_{U\in \mathbb{F}_2^m\backslash \{0\}, HW(U)\le m_0} \alpha_U^* (X_j^*)^U
\end{equation}
Now, we recluster the power features based on this low order model. We calculate the mean of the leakage as $\mu_l^*=\mathrm{E}_j[l_j^*]$, and define the clusters $C_0^*=\{\mathbf{c}_j|l_j^*>\mu_l^*\}$ and $C_0^*=\{\mathbf{c}_j|l_j^*<\mu_l^*\}$. The rank of key candidate $k^*$ is thus
\begin{equation}
    \mathcal{R}(k^*) = \max \big|\mathrm{E}_{\mathbf{c}_j\in C_0^*}[\mathbf{c}_j] - \mathrm{E}_{\mathbf{c}_j\in C_1^*}[\mathbf{c}_j] \big|
\end{equation}

\section{SCARL Attack on Ascon} \label{sec:ascon}
\begin{figure}
	\centering
    \includegraphics[width=\textwidth]{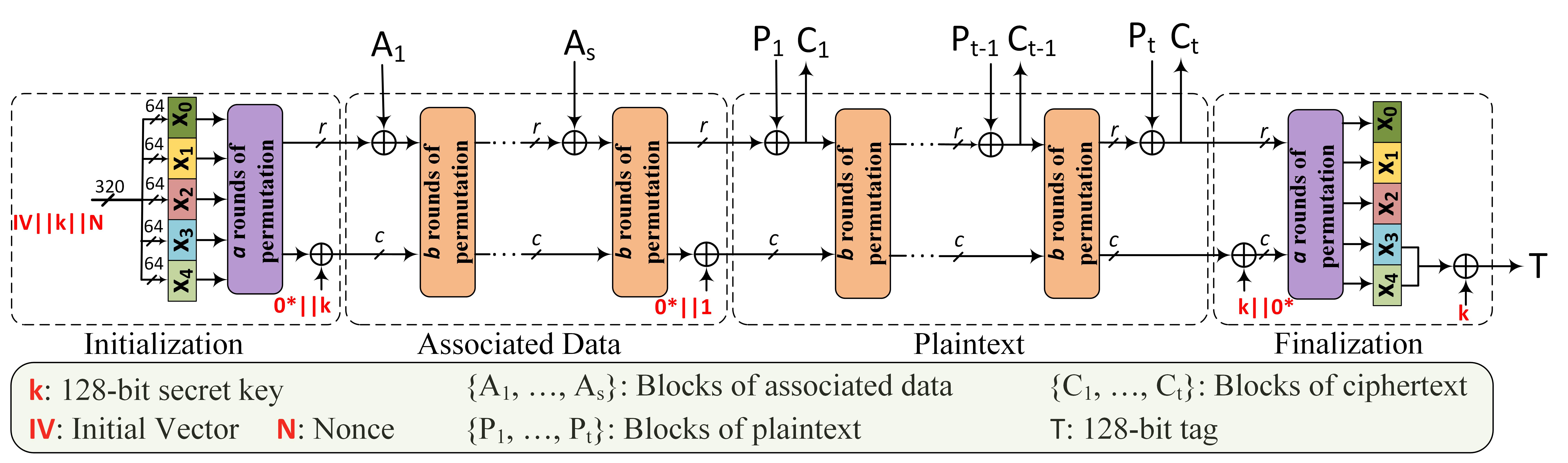}
	\vspace{-0.6cm}
	\caption{The Sponge structure of Ascon authenticated cipher with four stages of key insertion (Initialization), associated data, plaintext and tag generation \cite{ramezanpour2019statistical}}
	\label{fig:sponge}
\end{figure}

The Ascon authenticated cipher employs the so-called \textit{sponge} construction as shown in Fig.~\ref{fig:sponge} \cite{dobraunig2016ascon}. The internal 320-bit secret state of Ascon is arranged into five words of length 64 bits, denoted by $x_0$, $x_1$, $x_2$, $x_3$ and $x_4$. In a sponge construction, usually, the first $r$ bits of the state are called the rate bits and the remaining $c=320-r$ bits are the capacity. The $r$-bit blocks of associated data and the plaintext are added to the rate bits and the blocks of ciphertext are squeezed out of the same bits of the state. The unexposed capacity bits are mixed with the rate bits, to increase the entropy, during several rounds of permutation. The capacity of a sponge-based cipher is proportional to the privacy and authentication security bounds \cite{saarinen2014beyond}. 

To run an authenticated encryption for a plaintext with associated data, the state of Ascon is initialized with a 128-bit nonce, the 128-bit secret key and a 64-bit fixed initial vector (IV). Next, $a$ rounds of permutations are conducted on the state in the Initialization stage. The permutation function consists of the non-linear S-box operation, on a 5-bit vertical slice of the state, followed by the linear diffusion function, that combines the bits within a word of the state. The flow of the permutation function in the first round of the Initialization stage is shown in Fig. \ref{fig:power}. The first state word $x_0$ is initialized with the fixed IV. The bits of secret key $K=(k_0,k_1,\cdots,k_{127})$ are loaded into the next two words of the state, and the bits of nonce $n=(n_0,n_1,\cdots,n_{127})$ initialize the last two words.

\begin{figure}
	\centering
    \includegraphics[width=\textwidth]{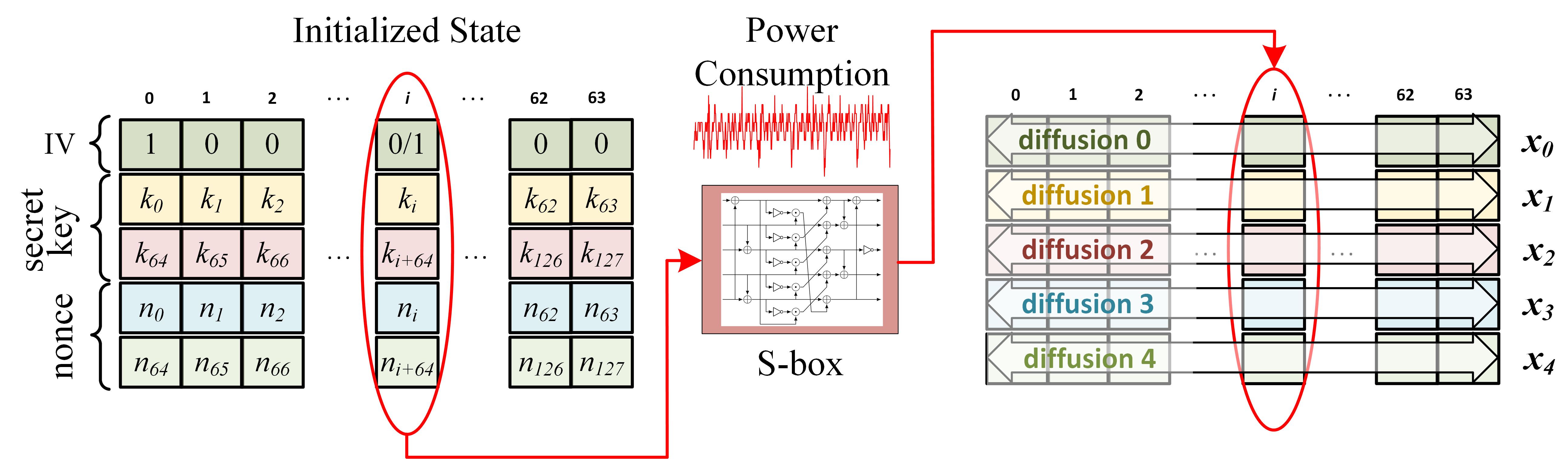}
	\vspace{-0.5cm}
	\caption{S-box operation and diffusion over the 320-bit state initialized with nonce, secret key and IV, at the beginning of Initialization stage of Ascon. The power consumption during S-box computations is used in the SCA attack.}
	\label{fig:power}
\end{figure}

\subsection{Attack Implementation} \label{sec:setup}
In the context of the attack model described in Section \ref{sec:clusterdiff}, the 5-bit S-box operations of Ascon, at the beginning of the Initialization stage, are the targets of our PA attack. At the beginning of Initialization, the most significant bit (MSB) at the input of the $i$-th S-box is the $i$-th bit of the IV which is known. The next two bits of the input are bits $k_i$ and $k_{i+64}$ of the secret key, respectively. The two least significant bits (LSB) of the input are bits $n_i$ and $n_{i+64}$ of the \textit{public message number}, or \textit{nonce}. We measure the power consumption of the S-box computation for multiple values of nonce. The known input data to the S-box $i$ is a 2-bit variable $Z_i\in\mathbb{F}_2^2=(n_i,n_{i+64})$ corresponding to two bits of the nonce. The secret data $X_i\in\mathbb{F}_2^5$ is the 5-bit variable at the output of the S-box. The subset of the key that is estimated using the power consumption of S-box $i$ is $k=(k_i,k_{i+64})$.

We implemented the Ascon authenticated cipher on an Artix-7 FPGA; the implementation is available for inspection at \cite{ascon}. The implementation is lightweight as it includes only one instantiation of the S-box hardware shared by all 64 S-box operations in Ascon. The S-box hardware is a bit-sliced implementation. After initializing the 320-bit state, the S-box operations in a round of permutation are conducted during the next 64 clock cycles, with one operation at every cycle. This implementation is consistent with the flow of operations shown in Fig. \ref{fig:power}. We measure the power consumption during S-box computations using the Flexible Open-source workBench fOr Side-channel analysis (FOBOS) \cite{fobos}.

Our FOBOS instance include a NewAE CW305 Artix-7 FPGA which executes the hardware implementation of the Ascon. This FPGA board is the target of the PA attack. In addition, FOBOS uses Digilent Nexys A7 as a control board for synchronizing the target with a host PC. The control board receives the input data, for encryption/decryption, from the PC and generates the proper signals for the clock and configuration of the target board. The power consumption of the target board during multiple encryptions, with random nonce values, is measured using a PicoScope 5000 with 20 dB amplification and 125 samples per clock cycle. The clock of the measurements is synchronized with the clock of the control board. 

\subsection{SCARL Attack}
\begin{figure}
	\centering
    \includegraphics[width=0.6\textwidth]{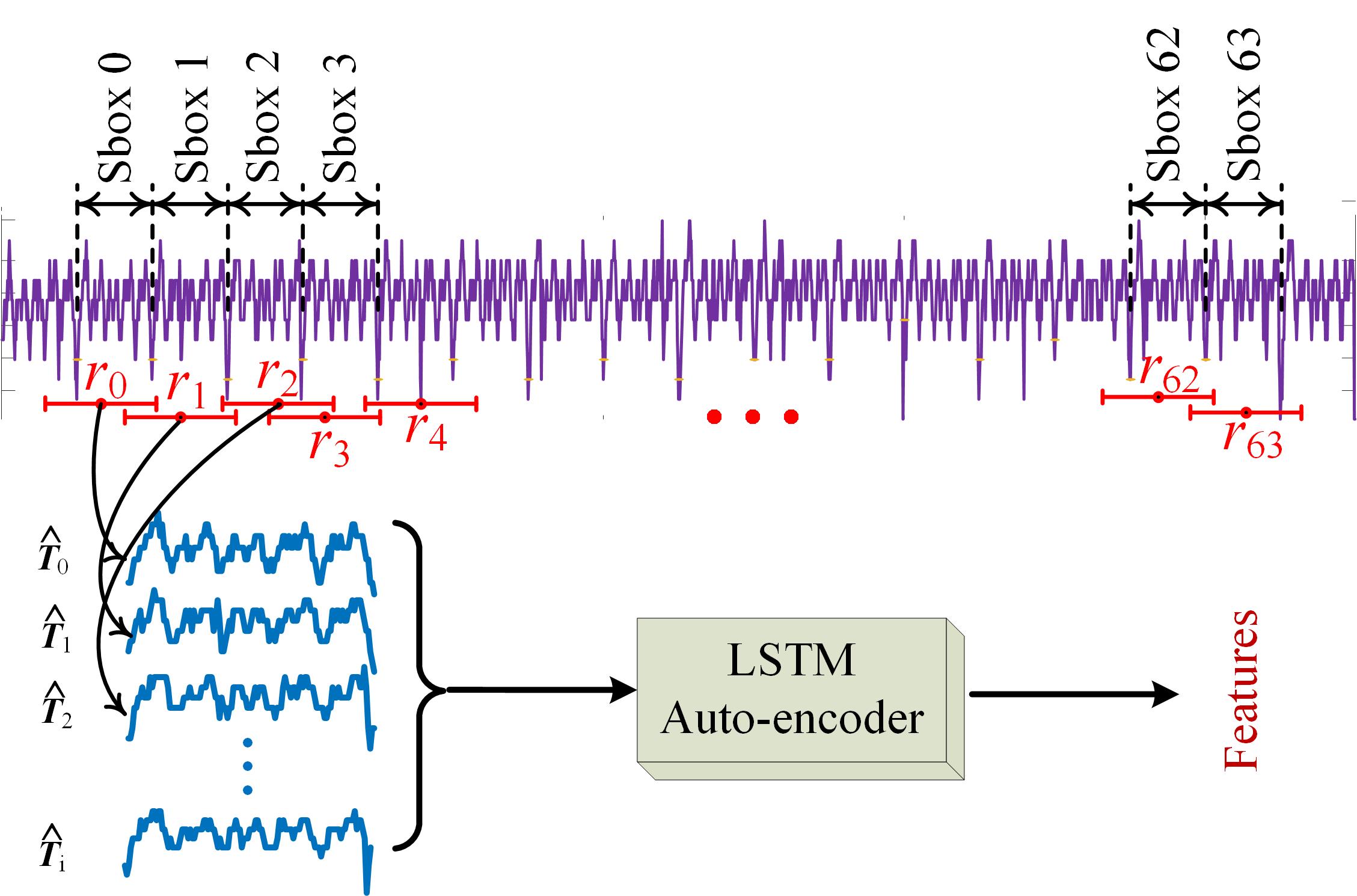}
	\vspace{-0.3cm}
	\caption{Horizontal processing of power measurements for separating power traces associated with individual S-boxes as the input to the autoencoder.}
	\label{fig:power}
\end{figure}

The first step in the SCARL attack is mapping the power measurements into an intermediate representation using the autoencoder. The power traces collected during the first round of Initialization, using FOBOS in our implementation of Ascon, include 64 times 125 consecutive samples corresponding to the power consumption of 64 S-box computations. If we know the rough timing of S-box operations, we can separate the power traces corresponding to the individual S-box operations in the entire power collected during round one of Initialization. 

We take $l$ samples of the power measurements centered at time instant $r_i$ as the power trace corresponding to the $i$-th S-box. This is shown in Fig. \ref{fig:power}. As mentioned earlier, each S-box computation is carried out in one clock cycle and 125 power samples are collected per cycle. Hence, the power traces of individual S-boxes has a size of 125 samples. However, we take the window length $l=125+\Delta l$ to address the uncertainty in locating the time of a S-box operation. We emphasize that the resolution of FOBOS is sufficiently high to accurately identify the time of all operations of the cipher. However, to demonstrate the power of SCARL in extracting data-dependent features, even with misaligned traces, we add the uncertainty $\Delta l$ in locating the power traces of individual S-boxes.

All power traces of an S-box, measured during multiple encryptions, can be used at the input of the LSTM autoencoder, described in Section \ref{sec:lastmauto}, to obtain the corresponding features. However, it is more efficient to use collective power traces of multiple S-boxes, as the size of data has a regularization effect in the learning process of the neural network. Assume we collect power measurements for $N$ encryptions. If we use the power traces of 2 S-boxes collectively, the autoencoder will be trained with a data set of size $2N$, i.e. twice the number of encryptions. Although the measurements of more than 2 S-boxes can be used collectively, due to memory limitations of the program running the neural networks, we use the power traces of pairs of S-boxes for extracting features by the autoencoder.

\begin{figure}
	\centering
    \includegraphics[width=\textwidth]{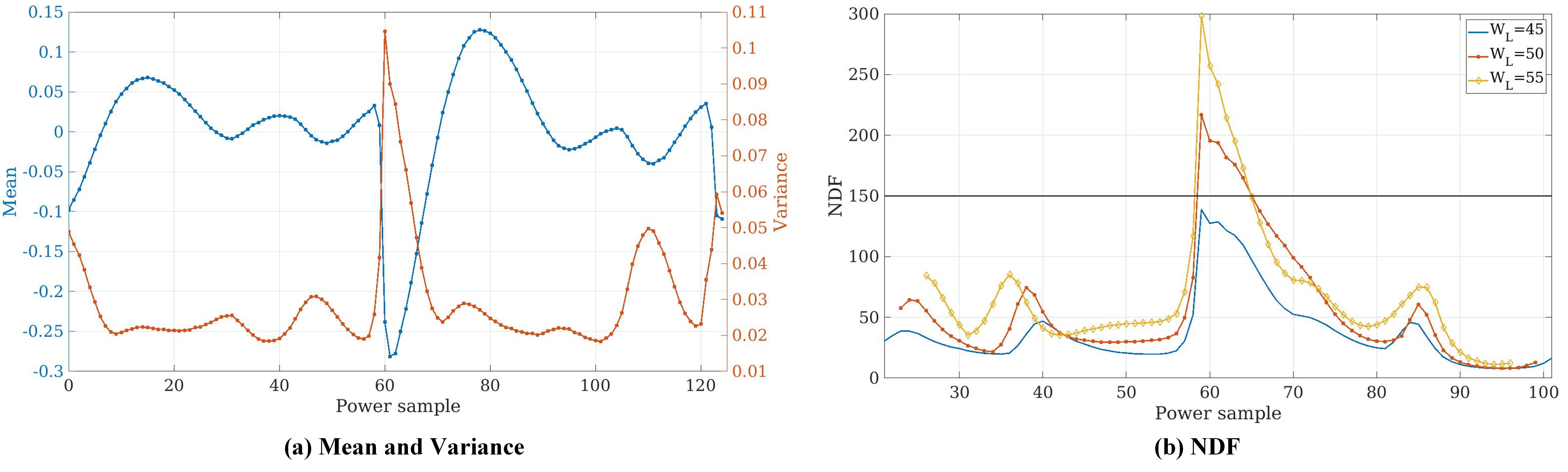}
	\vspace{-0.8cm}
	\caption{Statistical moments of power traces during S-box computations of Ascon; (a) mean and variance, and (b) number of degrees of freedom (NDF).}
	\label{fig:ndf}
\end{figure}

To map power traces into the autoencoder feature space, we need to tune the dimension of the features. The dimension is the number of neurons in the FC components of the LSTM cell in (\ref{fig:lstm_auto}). The mean and variance of the collective power traces corresponding to a pair of S-boxes for 30K encryptions is shown in Fig. \ref{fig:ndf} (a). To make an estimate for the number of degrees of freedom (NDF), we take a window of length $W_L$ around a power sample $i$. We calculate the spectrum of the covariance matrix of the samples in the window. According to the discussion of Section \ref{sec:ndf}, we find the NDF corresponding to sample $i$ as $W_L\sigma_i^2/\omega_0$, in which $\sigma_i^2$ is the variance of the sample and $\omega_0$ is the DC frequency component of the spectrum. The calculated NDF for different window lengths $W_L=45,50,55$ is shown in Fig. \ref{fig:ndf} (b). 

We set the dimension of the features in the LSTM autoencoder of Fig. \ref{fig:lstm_auto} to $D=150$. Hence, the NDF of the reconstructed power traces will be limited to 150. It is observed in Fig. \ref{fig:ndf} that The NDF of samples with higher variance increases beyond 150 when the window length around the samples is larger than $W_L=45$. This implies that the autoencoder smoothes variations with correlations times shorter than $45$ samples.

The features of the power traces, with dimension $D=150$, are clustered by the actor-critic network of \ref{fig:actocritic}. The inputs to the networks, thus, have a dimension of 150, as shown in the figure. A low order leakage model is estimated that fits the clusters for every key candidate, as described in Section \ref{sec:loworderleak}. The highest order of a generic leakage model for the 5-bit Ascon S-box is equal to 5. We limit the order of the estimated leakage model to 2. Hence, the estimated leakage model of (\ref{eq:lowmodel}) will be a second order model.

\section{Experimental Results} \label{sec:results}
We demonstrate the SCARL attack in recovering the secret key of the Ascon authenticated cipher from power consumption of the S-box operations at the beginning of the Initialization stage. The attack setup is described in Section \ref{sec:setup}. The power traces are measured at the supply pin of the target FPGA executing the Ascon implementation.

We implement the neural networks using Tensorflow on a PC with Intel Core-i7 CPU, 16 GB RAM, and Nvidia GeForce GTX 1080 GPU. The LSTM autoencoder takes 5 minutes to extract features of 60K power traces, which correspond to the power traces of a pair of S-boxes. The actor-critic networks for clustering the power features take 3 minutes. The rest of the SCARL algorithm, including low order leakage estimation and reclustering, for all key candidates, is completed in less than a minute for 60K traces, corresponding to a pair of S-boxes. Using two S-boxes, 4 bits of the secret key are unknown. As a result, the time required for SCARL to find 4 bits of the secret key is about 8 minutes.

\subsection{Classical DPA and CPA}
To compare the efficiency and power of a SCARL attack with existing SCA techniques, we also conducted DPA and CPA attacks on the Ascon implementation. In contrast to supervised learning SCA, DPA and CPA can be considered as unsupervised techniques if the attacker has a full knowledge of the leakage model. We demonstrate that SCARL significantly outperforms DPA and CPA attacks even though it does not require a leakage model.
\begin{figure}
	\centering
    \includegraphics[width=\textwidth]{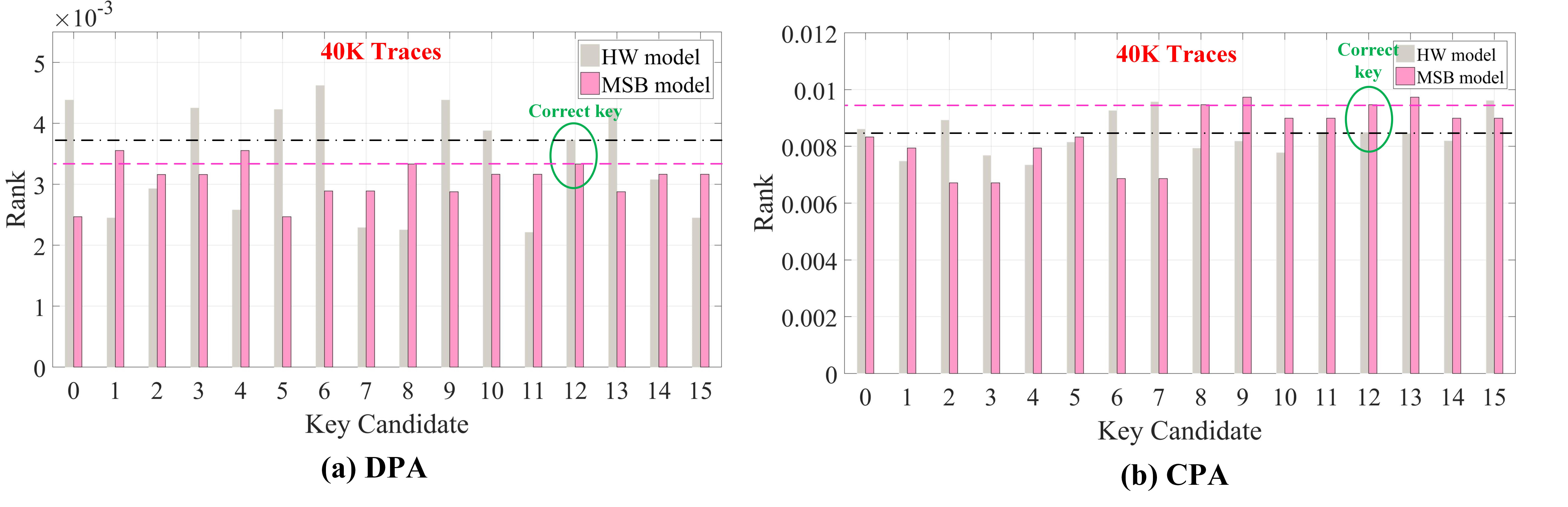}
	\vspace{-0.8cm}
	\caption{Statistical moments of power traces during S-box computations of Ascon; (a) mean and variance, and (b) number of degrees of freedom (NDF).}
	\label{fig:dpa}
\end{figure}

The DPA attack is conducted with two commonly used leakage models, i.e. Hamming weight (Hw) and most significant bit (MSB). In the Hw model, the leakage of secret data in (\ref{eq:leakmodel}) is $L^{Hw}(X)=\sum_i x_i$, in which $X=(x_0, x_1, \cdots, x_4)$ are the 5-bit values at the output of the S-boxes under attack. Using the MSB model, the leakage is simply $L^{MSB}=x_0$.
In a DPA attack, the power traces are grouped into two clusters based on the leakage. 

We collected power measurements of the FPGA for 40K encryptions with random nonce values. Let $\mathbf{T}_j$ denote the power measurements during encryption with the corresponding nonce value $n_j=(n_{j,0}, n_{j,1}, n_{j,2},\cdots,n_{j,127})$ for $j=1,2,\cdots,40K$. For the DPA attack, we separated the power traces of S-boxes 0 and 1. Hence, a total of 80K power traces $\mathbf{\hat T}_j, j=1,2,\cdots,80K$ are available, in which the first 40K traces correspond to S-box 0 and the rest to S-box 1. The corresponding input data are denoted by $Z_j=(n_{j,0}, n_{j,64}), j=1,2,\cdots,40K$ and $Z_j=(n_{j,1}, n_{j,65}), j=40K+1,40K+2,\cdots,80K$. The subset of the secret key (key candidate) that is estimated using this attack is $k=(k_0,k_{64},k_{1},k_{64})$, with 16 possible values.

For every value of the key candidate $k=k^*$, we calculate the output of the S-boxes. The output values $X_j^*,j=1,2,\cdots,80K$ are calculated with $k^*$ and the corresponding $Z_j$. The values of data leakage are then calculated as $L^{\mathcal{G}}(X_j^*)$, in which $\mathcal{G}$ is either Hw or MSB. Let $\mu_L^*$ denote the mean of the calculated leakage values. Next, we assign power traces with $L^{\mathcal{G}}(X_j^*)<\mu_L^*$ to cluster $C_0^*$ and traces with $L^{\mathcal{G}}(X_j^*)>\mu_L^*$ to cluster $C_1^*$. Finally, we calculate the difference between the mean traces in the two clusters. The sample with the highest difference is the rank of key candidate $k^*$.

The result of the DPA attack using Hw and MSB leakage models is shown in Fig. \ref{fig:dpa} (a). It is observed that with the Hw leakage model there are 7 incorrect key candidates with a higher rank than the correct key. However, with the MSB model, the rank of the correct key is among the top 4 key candidates. We note that rank of the correct key and the key candidate 8 are equal in this case. These results imply that the power consumption of S-boxes in Ascon are more correlated with the MSB of the output values.

The CPA attack is similar to DPA but with different rank statistics. Given the power traces $\mathbf{\hat T}_j, j=1,2,\cdots,80K$ and the calculated leakage values $L^{\mathcal{G}}(X_j^*)>\mu_L^*$, as described above, the Pearson's correlation coefficient between the samples of the power traces and the leakage values is used to rank the key candidate $k^*$. The sample with the highest absolute value of the correlation coefficient determines the rank of the key. The results of the CPA attack are shown in part (b) of Fig. \ref{fig:dpa}. Similar to DPA, the MSB leakage model shows improved performance. However, there are still two incorrect key candidates with higher rank than the correct key with 40K power measurements. Again, the incorrect key 8 has the same rank as the correct key.

\subsection{Results of SCARL}
The first step in SCARL attack is to map the raw power traces into an intermediate representation that captures the information content of all samples in the measurements. We use the LSTM autoencoder of Fig. \ref{fig:lstm_auto}, with 150 neurons in the FC components of the encoder/decoder cells, for this purpose. To find the optimal weights of the neural networks, we use the Adaptive moment estimation (Adam) algorithm for updating the weights using the MSE objective, described in Section \ref{sec:autoecnoder}. 
We use the batch normalization method with a batch size of 512. 

\begin{figure}
	\centering
    \includegraphics[width=\textwidth]{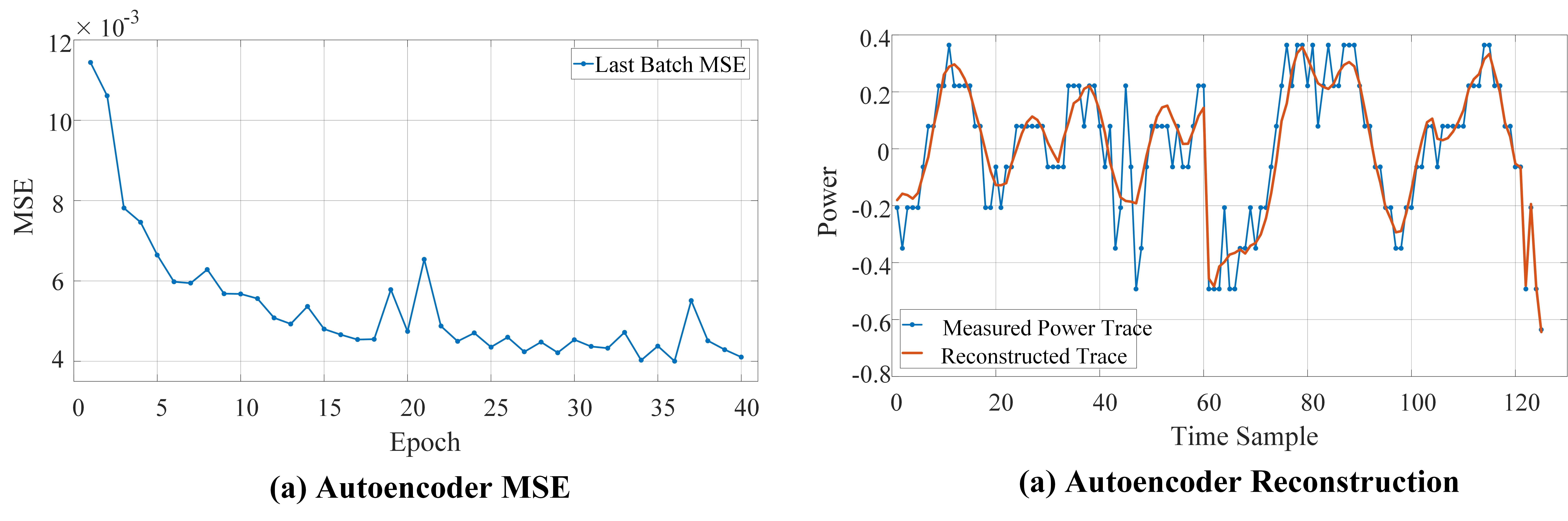}
	\vspace{-0.8cm}
	\caption{Performance of LSTM autoencoder in mapping measured power traces into the internal feature space; (a) MSE versus epochs of learning, and (b) measured power and reconstructed trace by the autoencoder.}
	\label{fig:auto_power}
\end{figure}

The raw power traces at the input of the autoencoder are normalized such that all samples of the traces are within $[-1,1]$. For a power trace $\mathbf{\hat T}=\{t_i|i=0,1,\cdots,N\}$, the normalized trace $\mathbf{\Tilde{T}}$ is obtained by subtracting the mean $\mathrm{E}_i[t_i]$ from $\mathbf{T}$ and then dividing the result by $\max_i(t_i)-\min_i(t_i)$. As shown in Fig. \ref{fig:lstm_auto}, the normalized power traces are first processed by a sliding window. We select windows of length $W=10$ and stride $S=5$. The 10 samples in $t$-th window are the input to the encoder cell at time instant $t$. The MSE of the autoencoder during successive epochs of learning is shown in Fig. \ref{fig:auto_power} (a). 

The MSE of Fig. \ref{fig:auto_power} (a) corresponds to the last batch of data at every learning epoch. It is observed that the variation of MSE is almost stabilized after 22 epochs. We stop the learning process at epoch 25. The reconstructed power trace, at the output of the decoder, for one measured power trace is shown in part (b) of the figure. We observe that the autoencoder identifies the major features of the power traces while high frequency variations, i.e. with lower correlation time, are filtered. 

\begin{figure}
	\centering
    \includegraphics[width=0.6\textwidth]{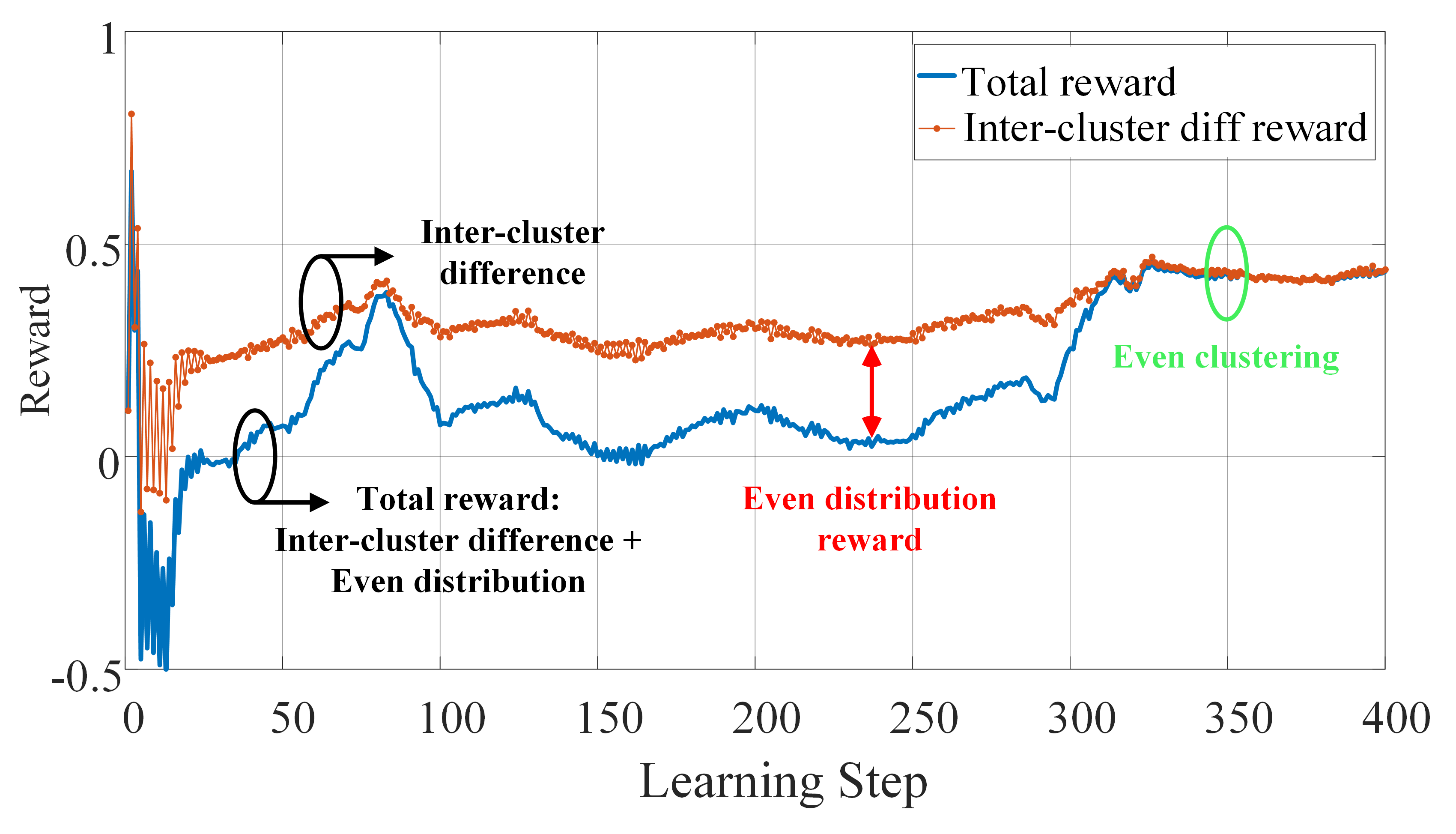}
	\vspace{-0.5cm}
	\caption{Obtained reward during successive steps of learning with the actor-critic networks and TD advantage learning for clustering power features.}
	\label{fig:reward}
\end{figure}

The extracted power features by the autoencoder are clustered by the actor-critic networks of Fig. \ref{fig:actocritic} based on the rewards defined in Section \ref{sec:actor-critic}. According to (\ref{eq:reward}), the two components of the reward measure the mean inter-cluster difference of the features in two clusters and the even assignment of features to the clusters. The total reward and the mean inter-cluster difference of features at successive steps of learning are shown in Fig. \ref{fig:reward}. The difference between these two curves is the reward component measuring the even assignment of clusters.

At the initial steps of learning the total reward is negative with large variations, as shown in Fig. \ref{fig:reward}. This means that the assignment of power features to the clusters is not even. Further, the inter-cluster difference is low which implies that the actor has not learned the pronounced features. When learning proceeds, the inter-cluster difference reward improves. Although this reward is almost stable during intermediate learning steps, the total reward is much smaller than the inter-cluster reward until step 300. This implies that the features are not evenly distributed over different clusters. After step 300, total reward is almost equal to the inter-cluster difference, i.e. the cluster assignment is now even, while the rewards are also stable. We stop the learning process at step 350.

\begin{figure}
	\centering
    \includegraphics[width=\textwidth]{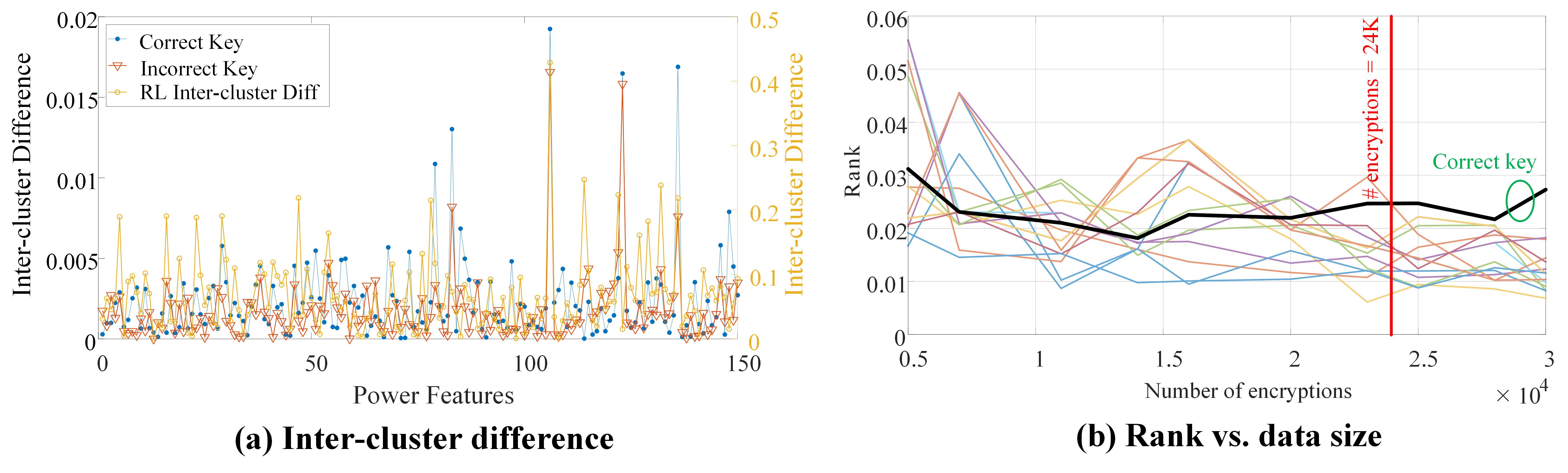}
	\vspace{-0.8cm}
	\caption{Rank of key candidates in SCARL attack; (a) inter-cluster difference of power features with the correct key and the incorrect key with highest rank; (b) rank of key candidates versus the number of power measurements.}
	\label{fig:rank_vs_size}
\end{figure}

After clustering the power features, a low order leakage model is estimated for every key candidate, as described in Section \ref{sec:loworderleak}. The power features are reclustered with this leakage model. The maximum mean difference of the features between two clusters is our rank statistics for the key candidates. The mean difference of power features, clustered with the low order leakage model, for the correct key and the incorrect key with the highest rank, for 30K measurements, is shown in Fig. \ref{fig:rank_vs_size} (a). In the same figure (right axis), the mean difference of features in clusters as found by the reinforcement learning algorithm, is also shown. We note that the maximum feature of the correct key coincides with the maximum inter-cluster difference obtained by RL. It implies that the RL algorithm has identified the information about the secret data from the power measurements.

The rank of key candidates versus the number of power measurements (encryptions) is also shown in Fig. \ref{fig:rank_vs_size} (b). If at least 24K measurements are collected, SCARL is able to identify the correct key. In this figure, the rank of key candidate corresponding to 4 bits of the secret key at the input of S-boxes 0 and 1 is plotted. This is in contrast to the results of DPA and CPA attacks, in Fig. \ref{fig:dpa}, which fail to identify the correct key with 40K measurements.

\section{Conclusions} \label{sec:conclude}
We demonstrated the application of deep reinforcement learning to develop a method of deep learning side-channel analysis, called SCARL, that extracts information about the secret data in a self-supervised approach. SCARL encodes the information content of all samples of power measurements into the internal representation of an autoencoder. We employed reinforcement learning, using actor-critic networks, to cluster the power features. Assuming that lower order terms of a generic leakage model have higher data leakage, SCARL can identify the proper leakage model from the clusters. We also demonstrated that SCARL is highly efficient compared to model-based attacks. With a lightweight implementation of the Ascon authenticated cipher on the Artix-7 FPGA, SCARL is able to recover the secret key using power measurements during 24K encryptions, while DPA and CPA attacks, with Hw and MSB leakage models, were unable to recover the key with 40K encryptions.

\bibliographystyle{ACM-Reference-Format}
\bibliography{bibliography.bib}


\end{document}